\documentclass[fleqn,usenatbib]{mnras}

\usepackage{newtxtext,newtxmath}

\usepackage[T1]{fontenc}

\DeclareRobustCommand{\VAN}[3]{#2}
\let\VANthebibliography\thebibliography
\def\thebibliography{\DeclareRobustCommand{\VAN}[3]{##3}\VANthebibliography}

\usepackage{graphicx}	
\usepackage{amsmath}	
\graphicspath{{figures/}}
\usepackage{url}
\usepackage{multirow}

\newcommand{\msun}{M$_\odot$}
\newcommand{\msunpcsq}{M$_\odot$ pc$^{-2}$}

\newcommand{\gureft}{\textsc{gureft}}

\defcitealias{Yung2024}{Y24}

\usepackage{xcolor}
\usepackage{soul}

\title[Density modulated star formation efficiency]{Density modulated star formation efficiency: implications for the observed abundance of ultra-violet luminous galaxies at $z>10$}

\author[R. S. Somerville et al.]{Rachel S. Somerville$^{1}$\thanks{e-mail: rsomerville@flatironinstitute.org}
L. Y. Aaron Yung,$^{2}$
Lachlan Lancaster,$^{3,1}$\thanks{Simons Fellow}
Shyam Menon,$^{4,1}$
Laura Sommovigo,$^{1}$ \newauthor
Steven L.\ Finkelstein$^{5}$
\\
$^{1}$Center for Computational Astrophysics, Flatiron Institute, 162 5th Ave, New York, NY 10010, USA\\
$^{2}$Space Telescope Science Institute, 3700 San Martin Drive, Baltimore, MD 21218, USA\\
$^{3}$Department of Astronomy, Columbia University, 550 W 120th St, New York, NY 10025, USA\\
$^{4}$Department of Physics and Astronomy, Rutgers University, 136 Frelinghuysen Road, Piscataway, NJ 08854, USA\\
$^{5}$Department of Astronomy, The University of Texas at Austin, Austin, TX, USA
}

\date{Accepted XXX. Received YYY; in original form ZZZ}

\pubyear{\the\year{}}

\begin{document}
\label{firstpage}
\pagerange{\pageref{firstpage}--\pageref{lastpage}}
\maketitle

\begin{abstract}
The number density of UV luminous galaxies discovered by the James Webb Space Telescope at ultra high redshift ($z \gtrsim 10$) is higher, and declines much more slowly with increasing redshift, than expected from extrapolations of lower redshift observations or pre-launch physics-based models. Most of these models assume star formation efficiencies (SFE) of only a few percent, motivated by observations of nearby galaxies. In this work, we incorporate a scaling of SFE with gas surface density (which we refer to as Density Modulated SFE; DMSFE), motivated by cloud-scale simulations and theory, into a semi-analytic cosmological model (SAM) of galaxy formation which is calibrated to match the observed rest-UV sizes of high redshift galaxies. We also model the impact of dust and bursty star formation on the SAM-predicted properties of observed galaxies. We show that with plausible values of the main parameters, such as the fraction of gas in dense clouds $f_{\rm dense}$, our new models easily reproduce or even exceed the observed galaxy number densities at $z\sim 6$--17. While no single value of $f_{\rm dense}$ is able to reproduce the very shallow observed decline of the galaxy number density at $z\gtrsim 12$, it is plausible and even expected for $f_{\rm dense}$ to have some effective dependence on cosmic time, which could bring these models into closer agreement with the data. We show that the combined effects of DMSFE, decreasing dust attenuation, and increasingly bursty star formation at earlier cosmic epochs could conspire to reproduce the observed evolution. 
\end{abstract}



\section{Introduction}
\label{sec:intro}
In the standard theoretical paradigm for galaxy formation, based on the hierarchical structure formation picture arising from the $\Lambda$ Cold Dark Matter ($\Lambda$CDM) cosmological model, galaxies form within gravitationally bound dark matter dominated halos, which grow from density fluctuations seeded by inflation shortly after the Big Bang \citep{Blumenthal1984,White1978}. Within these dense, gravitationally bound objects, gas is able to cool efficiently and eventually to form stars. The first stars are thought to form out of metal-free gas via H$_2$ cooling in mini-halos with masses of $10^5$--10$^6$ \msun\ \citep[][and references therein]{Klessen2023}, with more efficient atomic cooling arising as more massive objects, with temperatures $T\gtrsim 10^4$ K (masses of a few $\times\, 10^7$--$10^8$ \msun) begin to collapse. Within the $\Lambda$CDM paradigm, the number density of halos of a given mass at a given cosmic epoch can be predicted with great certainty under a specific set of assumptions about the nature of the dark matter and the cosmological parameters, using modern $N$-body simulations \citep[e.g.][]{Bagla2005,Nagamine2018,Klypin2011,Klypin2016}. The biggest open questions pertain to how the properties of these dark matter halos are related to observable galaxy properties, and how baryonic processes such as cooling, star formation, and stellar and black hole feedback shape these relationships. 

Numerically simulating these baryonic processes is a much more difficult problem than simulating the dark matter component alone, as the key processes that shape galaxies depend on physics occur across a vast range of scales, from less than a parsec to tens of Mpc. These processes include the accretion of gas from the Intergalactic Medium (IGM) into the circumgalactic medium (CGM), the cooling of gas from the CGM into the denser interstellar medium (ISM), collapse of warm neutral interstellar gas to form Giant Molecular Clouds (GMC), formation of dense GMC cores, and eventually, stars. Once massive stars ignite, they dramatically impact their surroundings via stellar winds and proto-stellar jets, photo-ionizing radiation and radiation pressure on dust \citep{Kimjg2018,Grudic2022}.  When the massive stars born in GMCs die as supernovae, they inject energy and momentum into the surrounding ISM, driving both galaxy scale outflows and turbulence which regulates star formation on larger scales \citep{Hopkins2012,Kimcg2017}. 

It is currently not computationally feasible to explicitly model all of these processes numerically, and thus all galaxy-scale and cosmological simulations adopt, to one extent or another, so-called ``sub-grid'' recipes to treat processes below the scale that can be explicitly simulated \citep{SomervilleDave2015,NaabOstriker2017}.  
In most existing simulations, these sub-grid recipes are somewhat \emph{ad hoc}, and are typically tuned empirically to match global galaxy observations in the nearby Universe, such as the stellar mass function of $z\sim 0$ galaxies. It is well known from observational and empirical studies that star formation in the Local Universe is strikingly \emph{inefficient} from GMC to halo scales. The star formation efficiency on the scales of GMCs to kpc scales is only 1-2 \% per free fall time \citep{Krumholz2007,Krumholz2019,Sunphangs2023}. Moreover, abundance matching analyses have shown that in the lower redshift Universe ($z \lesssim 6$), dark matter halos are only able to turn about 1-20\% of their universal ``budget'' of baryons ($f_b M_{\rm h}$, where $f_b$ is the universal baryon fraction and $M_h$ is the halo mass) into stars \citep{Moster2010,Behroozi2013c,Moster2018,Behroozi2019,Shuntov2022}. Most existing numerical simulations have tuned their sub-grid recipes for star formation and stellar feedback to match these low efficiencies. 

An alternative technique to fully numerical cosmological simulations for self-consistently simulating the complex, non-linear baryonic physics of galaxy formation is semi-analytic modeling \citep[see][for a review]{SomervilleDave2015}. In this approach, instead of explicitly solving the partial differential equations describing thermo- and hydrodynamics as well as gravity, coupled with the suite of sub-grid prescriptions, etc, one instead solves a system of ordinary differential equations (ODEs) that describe flows between different reservoirs, such as the IGM, CGM, ISM, and stellar body of a galaxy. These models are set within the cosmological backbone of halo ``merger trees" that describe the hierarchical formation of dark matter halos via accretion and mergers under the $\Lambda$CDM paradigm, as described above. Although orders of magnitude faster than numerical hydrodynamic simulations, SAMs have been shown to produce qualitatively similar predictions and insights \citep{SomervilleDave2015}. Like the sub-grid recipes used in numerical simulations, the scaling laws used in the SAM systems of ODEs also contain tunable parameters, which are similarly calibrated using a suite of global galaxy observables. 

A robust theoretical model should have \emph{predictive power} -- i.e., it should be able to make predictions for quantities or regimes for which the model has not been calibrated. One powerful way to stress test models of galaxy formation is to confront the model predictions with observations of galaxies at earlier and earlier cosmic epochs. This has been demonstrated over the past few decades with the discovery and characterization of higher and higher redshift galaxies with the Hubble and Spitzer Space Telescopes as well as ground-based 8-10m class telescopes \citep[see reviews by][]{Dunlop2013,Stark2016,Robertson2022,Madau2014}. This work characterized the UV luminosity functions and many other properties of galaxies up to about $z\sim 9$--10 \citep[e.g.][]{Oesch2018,Bowler2020,Finkelstein2022}. Although many semi-analytic models and numerical hydrodynamic simulations were not calibrated using observations at $z\gtrsim 0$, most models made predictions that were in reasonably good qualitative agreement with $z\sim 0$--10 UV luminosity functions and stellar mass function estimates \citep[e.g.][and references therein]{Yung2019a,Yung2019b}.

The launch of the James Webb Space Telescope \citep[JWST;][]{Gardner2023} has recently opened up an exciting window onto an even earlier cosmic epoch at $z\gtrsim 10$, which holds stringent challenges for galaxy formation models. Almost from the moment that the first data from JWST began to be analyzed, reports of new galaxy candidates that broke the $z\sim 10$ redshift barrier began to appear in the literature \citep{FinkelsteinMaisie2022,Castellano2022,Harikane2022,Bouwens2023}. It quickly became apparent that JWST was turning up a much larger number of UV-luminous galaxies than had been expected from empirical extrapolations of $z\lesssim 9$ observations, or than was predicted by physics-based pre-launch models \citep{Harikane2022,Finkelstein2023}. Over the past two and a half years, the uncertainties on the number density of these ultra-high-redshift galaxies have decreased as instrument calibrations have improved, and as they have been observed in multiple widely separated fields, improving the statistics and reducing uncertainties due to field-to-field clustering \citep{Adams2022,Adams2024,Castellano2023,Donnan2022,Donnan2024,Finkelstein2023,Finkelstein2024,Leung2023,Robertson2024,Whitler2025}. Spectroscopic redshifts have been obtained for large fractions of the objects, mostly confirming the robustness of the photometrically selected samples \citep{ArrabalHaro2023,Curtis-Lake2023,Wang2023,Fujimoto2023,Harikane2024,Harikane2025,Kokorev2025}. Another way to frame the surprise that JWST has unveiled was that the comoving number density of UV-luminous galaxies $n_{\rm UV bright}$ at $z\gtrsim 9$ \emph{declines much less rapidly} with increasing redshift than was expected empirically or predicted by pre-launch models \citep{Finkelstein2023,Leung2023,Finkelstein2024,Adams2024}. The number density of massive dark matter halos ($M_h \simeq 10^{10}$--$10^{11}$ \msun) declines much more steeply over this time interval, by two orders of magnitude or more \citep{Yung2024a,Robertson2024}. Therefore, if the observations are robust, the mapping between emergent UV luminosity and dark matter halo mass at these cosmic epochs must diverge significantly from what physics-based or semi-empirical models calibrated at lower redshifts predict, \emph{and must evolve significantly with redshift}. 

A wide range of possible solutions to this conundrum have been proposed. One class of solutions proposes fundamental modifications to cosmological structure formation. Some suggested solutions in this category include Early (evolving) Dark Energy \citep{Menci2022,Menci2024}, or adding extra small scale power to the primordial power spectrum via a blue spectral index \citep{Parashari2023,Hirano2024}, primordial black holes \citep{Liu2022,Colazo2024} or cosmic strings \citep{Koehler2024}. In this paper, we focus on the other class of solutions, in which we retain the fundamental framework of `vanilla' $\Lambda$CDM and consider modifications to baryonic processes. 

These can be further sub-divided into two scenarios: 1) galaxies are \emph{more luminous} (for the same underlying SFR or mass of stars produced) than the previous models predicted or 2) star formation is \emph{more efficient} in these early galaxies (implying that more stars were formed over this period). In the first category, it has been suggested that the stellar initial mass function (IMF) in these early galaxies could be top-heavy (richer in massive stars), leading to more rest-UV light per unit stellar mass formed \citep{Inayoshi2022,Trinca2024,Yung2024b,Mauerhofer2025}, or there could be a contribution to the UV light from accreting black holes \citep{Inayoshi2022,Volonteri2023,Fujimotouncover2024,Trinca2024}. Another proposed mechanism is to temporarily brighten a subset of intrinsically less massive galaxies (populating lower mass halos) due to a short-lived burst of star formation \citep{Mason2023,Shen2023,Sun2023,Yung2024b,Kravtsov2024,Gelli2024}. 
Yet another phenomenon which could help explain the shallower-than-expected rise of UV-luminous galaxies with time is if dust attenuation in the UV increases rapidly over the relevant period. This could occur if dust is caused to expand or be ejected from galaxies by radiation pressure \citep{Ziparo2022,Ferrara2023,Ferrara2024}, if the dust grain size distribution was different in the past, leading to lower UV attenuation \citep{Narayanan2025}, or if early dust yields from SNe were lower, or dust was destroyed more efficiently, at these early epochs than generally assumed (though \citet{Ferrara2025} conclude that the latter is unlikely to be able to account for the observations). 

In the second category of scenarios that propose higher star formation efficiencies, \citet{Dekel2023} has pointed out that the ISM of these $z\gtrsim10$ galaxies is likely to be extremely dense and rather metal poor. If the density of individual star forming clouds is correspondingly denser than in the local universe, the associated cloud free fall times, $t_{\rm ff}$, are very short, $\lesssim$ 1 Myr. While \citet{Dekel2023} and \citet{Li2024} argue that high star formation efficiencies in this regime are the result of cloud free fall times being short compared to the time-scale at which SNe begin to explode ($t_{\rm ff} \ll t_{\rm SN}$), high resolution cloud-scale simulations indicate that a significant fraction of star formation occurs on time-scales shorter than $t_{\rm SN}$, even in much lower density clouds \citep{Kimjg2018,Lancaster2021,Grudic2022,Menon2023,Polak2024}. This body of work suggests that SNae are usually not important for determining the final star formation efficiency of GMCs, except perhaps in very low density environments. They are, of course, still important for driving galactic scale winds and turbulence on the scale of the galaxy overall, as noted above. Instead, models of star formation that are tested against these same simulations indicate that the high cloud-scale SFEs (defined as the fraction of gas converted to stars over the lifetime of the cloud) achieved in this high surface density regime ($\Sigma_{\rm cl} \simeq 10^3$ -- $10^4$ \msun pc$^{-2}$) are due to the failure of the ``early'' (pre-SN) feedback from massive stars in its fight against the gravitational collapse of the cloud \citep{Thompson2016,Grudic2018,Lancaster2021}. Specifically, when the rate of specific momentum injection ($\dot{p}/M_*$) exceeds the gravitational weight, the gas will become unbound and star formation will be (at least temporarily) halted. Thus, for a given momentum injection rate, denser clouds are more difficult to unbind and will be able to convert a larger fraction of gas into stars before being dispersed. As a result, cloud-scale SFEs in dense, high-pressure environments can be much higher than is typical in the nearby Universe, reaching 85-90\% in the densest clouds \citep{Chevance2023,MenonIMF2024,Menonfesc2024}. 

If galaxy disk sizes scale in proportion to the virial radius of their host halo (which has been shown empirically to be the case from $z\sim 0$--7; \citealp{Somervillesizev2008,Shibuya2015}), then we would expect the typical baryonic mass surface densities in high redshift $z\gtrsim 10$ galaxies to be two or more orders of magnitude higher than in nearby galaxies. This is supported by the sizes of galaxies observed by JWST at $z\sim 9$--13 \citep{Casey2024,Morishita2024}. If the surface densities of individual star forming clouds scale with the average overall gas surface density, then we would expect the stars in high redshift galaxies to form in much denser clouds than those that are typical in nearby galaxies.  This is also supported by observations of extremely compact star forming clumps, possibly bound proto-clusters in formation, that have been seen in lensed high-redshift galaxies at $z\sim 6$--10 \citep{Vanzella2023,Adamo2024,Mowla2024,Fujimotograpes2024}. Taken together, the higher expected SFE in dense environments, and the expected evolution towards higher density structures overall at earlier times in the $\Lambda$CDM paradigm, appear to point towards inevitably higher SFE in high redshift galaxies, as recently pointed out by \citet{Boylan-Kolchin2025}.

In this work, we combine the insights from cloud-scale simulations, suggesting that the conversion of gas into stars is more efficient at high gas surface density, with a well-established semi-analytic model of galaxy formation set within a $\Lambda$CDM cosmological framework, the Santa Cruz semi-anaytic model (SC SAM), to make predictions of statistical properties of the galaxy populations at $6 \lesssim z \lesssim 18$. We compare these predictions with observations from JWST. In addition, we investigate the effects of dust attenuation and enhanced bursty star formation within our modeling framework using approaches inspired by models in the literature \citep{Ferrara2024,Gelli2024}. 

The structure of the paper is as follows. In Section~\ref{sec:methods}, we present a brief overview of the physical ingredients in the baseline Santa Cruz semi-analytic models, our new ``cloud based'' model for star formation, and our implementations of dust and enhanced starbursts.  In Section~\ref{sec:results}, we present our main results. We discuss the implications of our results in Section~\ref{sec:discussion} and summarize and conclude in Section~\ref{sec:conclusions}. 

\section{Methods}
\label{sec:methods}
The Santa Cruz semi-analytic model \citep{Somerville1999,Somerville2001} is a well-established tool that has been used to make predictions for galaxy populations spanning a broad range of properties and redshifts \citep{Somerville2008,Somerville2012, Popping2014,Somerville2015,Popping2017,Yung2019a,Yung2019b,Yung2022,Somerville2021}. The main ingredients of the model as used in this work are described in \citet[][S08]{Somerville2008} and \citet{Somerville2015}, and we refer to those works for details. The parameter values used in this work are the same as those used in \citet[][hereafter Y24]{Yung2024b}. In the remainder of this section, we provide a brief summary of the suite of high-redshift optimized $N$-body simulations from which we extract our halo merger trees (\S\ref{sec:gureft}), the most important aspects of the baseline model (\S\ref{sec:scsam}), and the details of our new cloud-based model for star formation (\S\ref{sec:cloudmodel}). We also describe how we model dust attenuation and bursty star formation in post-processing, in \S\ref{sec:dust} and \S\ref{sec:burstmodel}.
Throughout this work, we adopt cosmological parameters $\Omega_m = 0.308$, $\Omega_\Lambda = 0.693$, $H_0 = 67.8$ km s$ ^{-1}$ Mpc$ ^{-1}$, $\sigma_8 = 0.823$, and $n_s = 0.96$; which are consistent with the ones reported by the Planck Collaboration (Planck Collaboration XIII \citeyear{Planck2016}).

\subsection{GUREFT merger tree suite}
\label{sec:gureft}
Gadget at Ultrahigh Redshift with Extra-Fine Timesteps (\gureft, pronounced \textit{graft}) is a recent suite of dark matter-only cosmological simulations, carried out with the publicly available version of the \textsc{Gadget}-2 code\footnote{\url{https://wwwmpa.mpa-garching.mpg.de/gadget/}} \citep{Springel2005}, that was specifically designed to accurately capture the merger histories of the dark matter halos that host galaxies at very high redshifts ($z\gtrsim 6$; \citealp{Yung2024a}). The suite contains four different periodic cubic volumes (5, 15, 35, and 90 Mpc\,$h^{-1}$ on a side), each simulated with $1024^3$ dark matter particles, resulting in DM particle masses from $1.5 \times 10^4$ \msun\ to $8.5 \times 10^7$ \msun, respectively. In this work, we supplement the \gureft\ suite with the Very Small Multidark Planck (VSMDPL) simulation from the Multidark suite \citep{Klypin2016}, which has a volume of (160$/h$)$^3$ Mpc and a particle mass of $9.1 \times 10^6$ \msun, and the same cosmological parameters as \gureft.
Results from these five boxes can be ``grafted'' together to cover a larger dynamic range than any currently available single simulation.  Unlike many previous $N$-body simulations, which saved only a few snapshots at $z\gtrsim 10$, \gureft\ stored 170 snapshots between $z = 40$ to 6 with a spacing roughly one-tenth of the typical halo dynamical time. Halo catalogs were generated using the \textsc{rockstar} halo finder \citep{Behroozi2013a}, and merger trees were constructed using the \textsc{consistent-tree} algorithm \citep{Behroozi2013b}. We adopt the halo virial mass and virial radius definitions corresponding to the definition of \citet{Bryan1998}. For more information on the \gureft\ suite as well as its predictions for halo mass functions, halo concentrations, halo spins, and halo mass accretion histories, along with convenient fitting functions, see \citet{Yung2024a,Yung2025}. 

\begin{figure*}    
    \includegraphics[width=2.08\columnwidth]{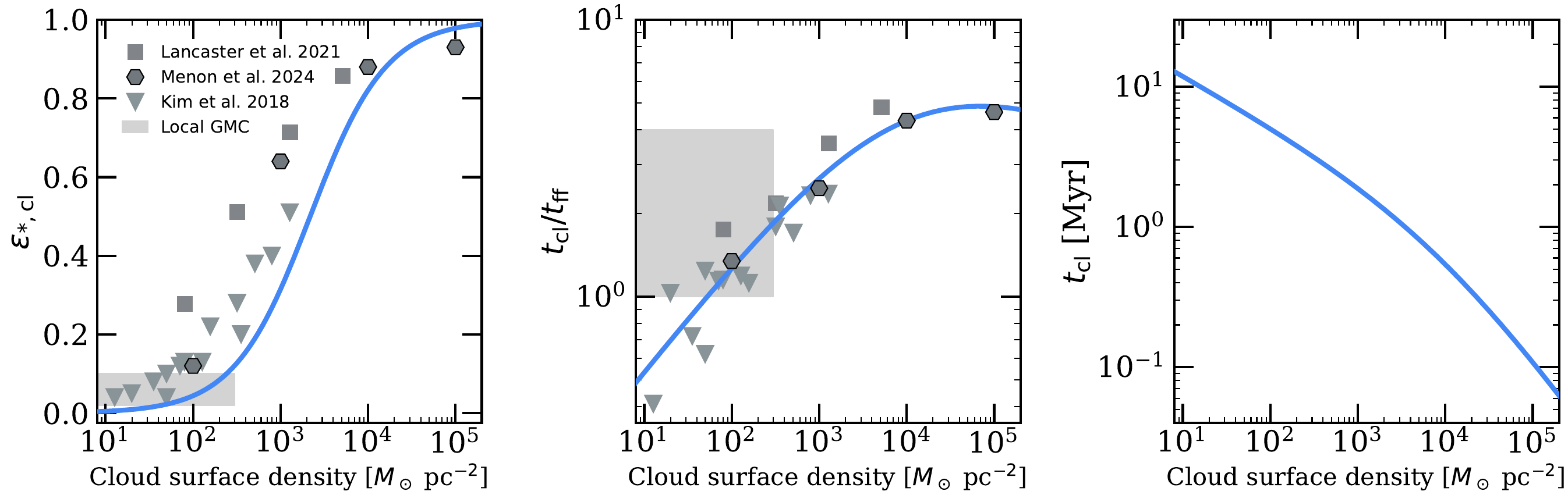}
    \caption{
        Star formation efficiency per star forming cloud (left) and cloud lifetimes (in units of the free fall time (middle), and in Myr (right)) as a function of cloud surface density. The grey shaded regions represent the star formation efficiency (left panel), lifetimes (middle), and surface density range of GMCs in local universe star forming galaxies \citep{Chevance2023}. Symbols show cloud-scale star formation efficiencies (integrated over the cloud lifetime; left) and lifetimes (middle) from cloud-scale simulations by \citet[][squares]{Lancaster2021}, \citet[][hexagons]{Menonfesc2024}, and \citet[][triangles]{Kimjg2018}. The solid blue lines (left and middle) show the analytic scalings for $\epsilon_{\rm *, cl}$ and cloud lifetime that are used in the SAM (see text).         
    }
    \label{fig:estar_tcloud}
\end{figure*}

\subsection{Baseline Santa Cruz semi-analytic model for galaxy formation}
\label{sec:scsam}

Within the halo merger trees from the \gureft\ suite described above, the Santa Cruz SAM (SC-SAM) solves a system of ordinary differential equations describing the flows of gas into and out of halos and galaxies, driven by cosmological accretion and cooling and supernova driven galaxy-scale outflows, the conversion of gas into stars, and the production and dispersal of heavy elements by stars and supernovae. 

A basic ansatz of SAMs, common to many empirical models, is that the rate that gas accretes into the CGM of a galaxy is proportional to the total growth rate of the halo, $\dot{M}_{\rm acc} = f_b \dot{M}_{\rm halo}$, where $f_b$ is the universal baryon fraction. This accretion rate is suppressed in low-mass halos by photoionization feedback after the universe is reionized \citep{Okamoto2008}. A simple cooling model, based on the cooling time defined as the time required for the gas to lose all of its thermal energy, determines how rapidly the CGM accretes into the ISM (S08). If the cooling time is shorter than the halo dynamical time (which is generally the case at the epochs of interest in this work), the gas is assumed to fall into the ISM on a dynamical time. 

Cooled gas is assumed to settle into a rotationally supported disk. In previous versions of the SC SAM, we have adopted a model based on the angular momentum partition ansatz \citep{Mo1998}, in which it is assumed that the halo gas acquires angular momentum along with the dark matter via tidal torques during the halo collapse, and that the specific angular momentum of the gas is conserved when it falls in to form a disk. In an isothermal halo, in the absence of self-gravity, this model predicts that the exponential scale radius of the disk $r_{\rm disk} \propto \lambda R_{\rm h}$, where $\lambda$ is the dimensionless spin parameter \citep{Peebles1969} and $R_{\rm h}$ is the virial radius of the halo. In previous works, we adopted a more detailed model for disk sizes, which incorporates the effects of a more realistic Navarro-Frenk-White halo profile, and the baryonic contraction of the gas due to its self-gravity \citep{Somervillesizev2008,Porter2014}. In this work, we adopt a simpler prescription in which the exponential scale radius of the disk $r_{\rm disk} = f_{r} \lambda R_{\rm h}$, where we adjust $f_{r}$ to match the observed sizes of high redshift galaxies (see \S\ref{sec:sizes}). We also assume that the radius of the cold, star forming ISM gas is the same as that of the stars, and we use $r_{\rm disk}$ interchangeably to refer to the radius of cold gas or stars. 

In the version of the Santa Cruz SAM used in many recent works (including \citealp{Yung2024b} and the Yung et al. JWST forecast series), the ISM was partitioned into atomic, molecular, and ionized components using fitting functions based on numerical hydrodynamic simulations by \citet{Gnedin2011}. The star formation rate density was then computed based on the molecular gas ($H_2$) density, using empirical scalings motivated by observations and theory \citep{Somerville2015}. However, the physical picture underlying the \citet{Gnedin2011} simulations as well as the notion that star formation traces $H_2$ may well break down at very low metallicities as shown by recent simulations \citep{Polzin2024,Semenov2025}. Therefore, in this work, for our baseline model we revert to a simpler star formation recipe used in the version of the SC-SAM presented by \citet{Somerville2008}, and is also similar to the star formation prescription used in many comological hydrodynamical simulations. Namely, we adopt a standard Kennicutt-Schmidt-like star formation law, in which the SFR surface density is proportional to the total cold ISM surface density:
\begin{equation}
\Sigma_{\rm SFR} = A_K \, \Sigma_{\rm gas}^{N_{\rm K}}
\end{equation}
where $A_K$ and $N_K$ are free parameters, and $\Sigma_{\rm gas}$ is the surface density of cold gas in the ISM. To compute the total SFR, this function is integrated above a critical surface density (see S08 for details). The critical gas surface density for star formation is supported by observations, and schematically represents the transition from predominantly atomic gas (H$_{\rm I}$ to molecular gas (H$_{2}$). See \citet{Somerville2015} for a more detailed discussion and comparison of the explicitly H$_{2}$-based SF model and the Kennicutt-Schmidt model. In addition, following galaxy mergers with a mass ratio greater than 0.1, a burst of star formation is triggered, as described in \citet{Somerville2008}. 

Outflows of mass and metals from the ISM to the CGM and IGM are modeled using a simple power law function of the maximum rotation velocity of the halo $V_{\rm max}$, parameterized by a normalization and a slope (S08). Gas that is ejected from the CGM is stored in a special reservoir, that is allowed to ``reaccrete'' back into the halo on a timescale $\chi_{\rm re-infall}\, t_{\rm dyn}$, where $\chi_{\rm re-infall}$ is a free parameter and $t_{\rm dyn}$ is the halo dynamical time. 

Our models also include a simple recipe for black hole seeding, black hole growth via accretion and mergers, and black hole feedback via winds and thermal heating of the CGM, but we have confirmed that, with the current modeling framework and parameters, black hole physics does not significantly impact any of the predictions shown in this work. However, we emphasize that more physically motivated models for black hole seeding could lead to more efficient early growth of black holes, and hence to greater sensitivity of our model predictions to black hole physics. We discuss this further in Section~\ref{sec:discussion}.

The free parameters in the model are tuned by hand to match a set of observation-based constraints at $z=0$, as shown in \citet{Yung2019a} and \citet{Gabrielpillai2022}, and are not adjusted to match any high redshift observations. 

To compute spectral energy distributions and mock photometry for our galaxies, we convolve the full star formation and chemical enrichment history with stellar population synthesis models. In this work we use the Binary Population and Spectral Synthesis (BPASS\footnote{\url{https://bpass.auckland.ac.nz/} v2.2.1}) models \citep{Eldridge2017, Byrne2022}, including binary stars. These model grids span (absolute) stellar metallicity values $Z = 10^{-5}$ to 0.040. The SSP models adopted in this work assume a fiducial broken power law stellar IMF, with an upper slope $a_1 = -1.30$ between $0.1$ -- $0.5$ \msun\ and a lower slope $a_2 = -2.35$ between $0.5$ -- $300$ \msun. Details of how the photometry is computed are described in Yung et al. (in prep). In this work, we include only stellar continuum emission, and neglect nebular emission, although we will include the latter in a forthcoming work. Throughout this work, where we refer to rest-UV, we adopt a tophat filter with a central wavelength of $1600$\AA. 

\begin{figure}
    \includegraphics[width=\columnwidth]{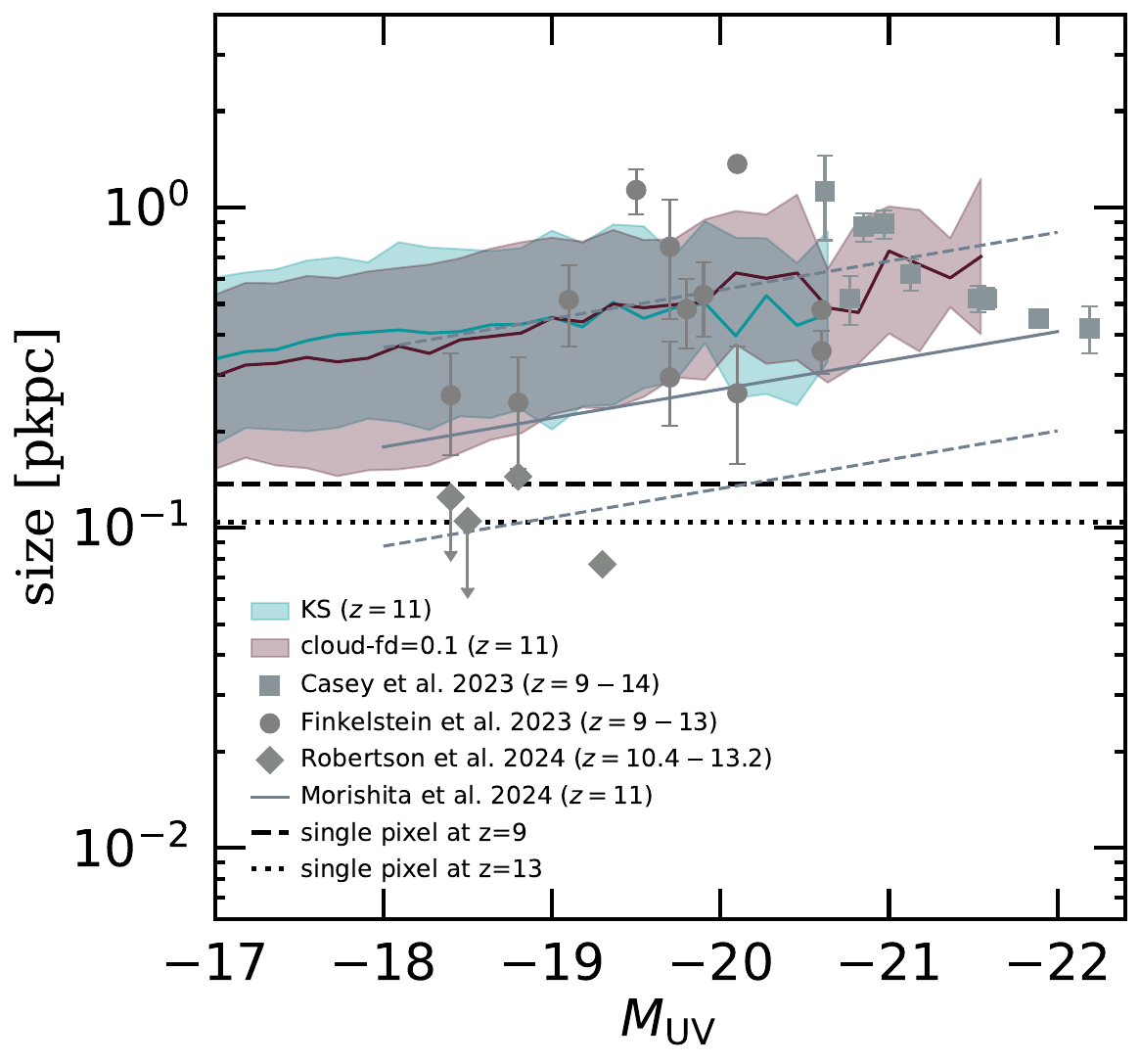}
    \caption{
    The (physical) half-light (or half-mass) radius of ultra-high redshift galaxies as a function of their rest-UV magnitude. Symbols show observed UV half-light radii for $9\lesssim z \lesssim 13$ galaxies from several recent JWST surveys, as indicated by the figure legend. The cyan and red lines show the median half-mass radii of galaxies in our \textbf{KS} and \textbf{cloud-fd=0.1} models at $z=11$, and shaded areas show the 16th and 84th percentiles. The solid grey line shows the fit to the median observed galaxy UV half-light radii at $z\sim 11$ from \citet{Morishita2024}, and the grey dashed lines show their quoted 1$\sigma$ dispersion around the median. The horizontal dashed and dotted lines show the physical size of one NIRCam pixel at $z=9$ and $z=13$, respectively. 
    }
    \label{fig:sizemag}
\end{figure}

\begin{figure}
    \includegraphics[width=\columnwidth]{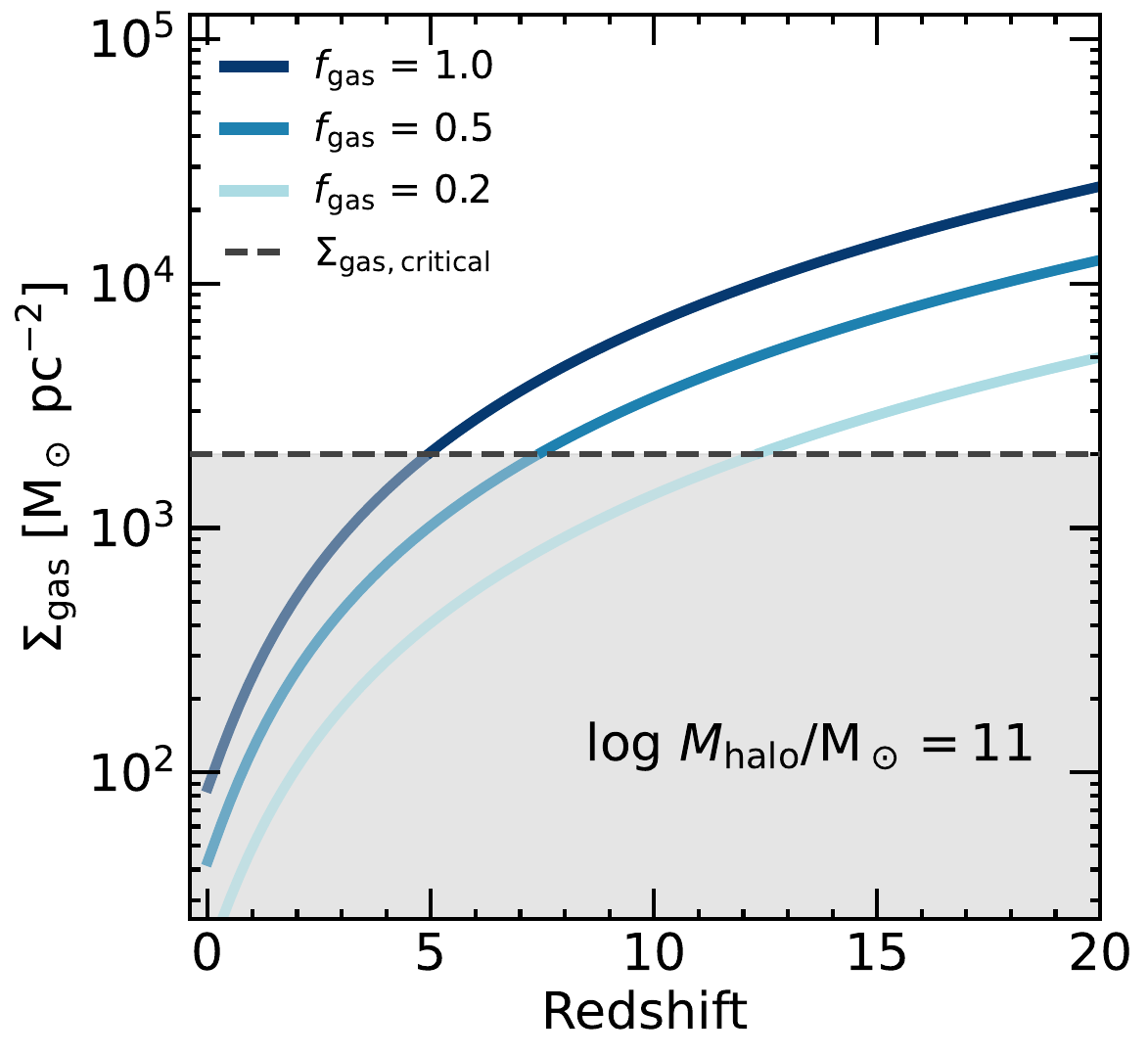}
    \caption{The gas surface density as a function of redshift for a simple toy example of a halo with mass $M_{\rm h}=10^{11}$ \msun\ at the plotted redshift, with a gas radius that is 0.06 times the halo virial radius, and a gas fraction $f_{\rm gas}=0.2$, 0.5, or 1.0, where $f_{\rm gas}=m_{\rm gas}/(f_b M_{\rm h})$ is the fraction of the halo baryon budget in the cold ISM. This illustrates that the gas surface density may have been up to two orders of magnitude higher in galaxies at ultra-high redshifts, and that galaxies may cross the critical gas surface density of $\Sigma_{\rm gas, critical} \sim 2000$ \msunpcsq, where stellar feedback begins to become ineffective, at around $z\sim 10$.
    }
    \label{fig:gasdensityz}
\end{figure}

\begin{figure*}
    \includegraphics[width=2\columnwidth]{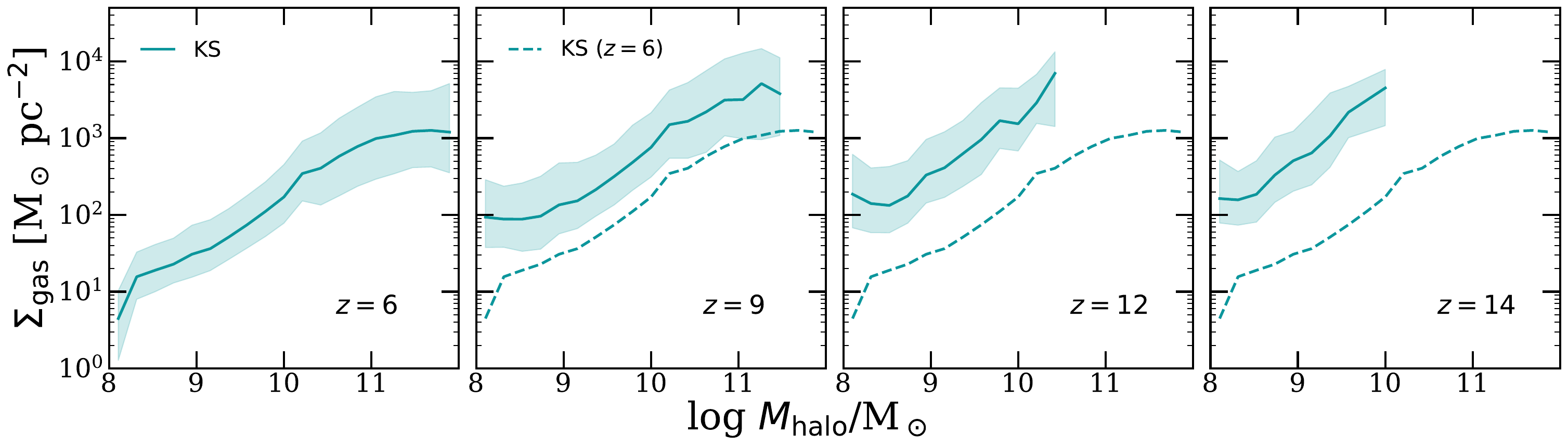}
    \caption{The average gas surface density of galaxies in our \textbf{KS} model as a function of halo mass, for several redshifts as marked on the panels. The solid line shows the median and the shaded region shows the 16th and 84th percentiles. The dashed line shows the relation at $z=6$, repeated in each panel to guide the eye. This figure shows that although the gas is predicted to be denser in early galaxies than in the local universe, the evolution is not quite as strong as that predicted by the simple toy model shown in Fig.~\ref{fig:gasdensityz}. This is because the gas fraction is not a constant in the SAM, but is determined by star formation and gas inflows and outflows, and shows trends with both halo mass and redshift. 
    }
    \label{fig:Sigma_Gas}
\end{figure*}

\subsection{Density modulated star formation model}
\label{sec:cloudmodel}
Our new model is motivated by results from simulations representing individual Giant Molecular Clouds (GMC) or proto-star-clusters that include detailed treatments of the primary feedback processes associated with massive stars, namely stellar winds and jets, ionizing and non-ionizing UV/optical radiation, and IR radiation from UV/optical radiation reprocessed by dust \citep[see review by][]{Chevance2023}. It is well known that in the local Universe, these processes disperse GMCs in just a few free fall times, yielding integrated star formation efficiencies (here defined as the fraction of the initial GMC gas mass that is turned into stars by the time the cloud is dispersed) of only a few percent \citep{Krumholz2007,Krumholz2019}. As shown in Fig.~4 of \citet{Chevance2023}, simulations from different groups and containing different physics all predict a strong trend in the integrated SFE with initial cloud surface density. For example, \citet{Lancaster2021} carried out GMC-scale simulations with stellar wind feedback in turbulent, self-gravitating clouds, probing surface densities up to $\sim 5000$ \msunpcsq. \citet{Kimjg2018} studied the impact of photo-ionization and radiation pressure on regulating cloud-scale SFE and lifetimes. \citet[][M24]{Menonfesc2024} used cloud-scale simulations with radiative transfer to simulate star formation in clouds with surface densities ranging from $\Sigma_{\rm cl} \sim 100$--$10^5$ \msunpcsq. All of these studies found that the cloud scale star formation efficiency (SFE), defined as the fraction of the initial mass of the cloud that has turned into stars by the end of the simulation, increases significantly with the initial cloud surface density, from values of $\epsilon_{*,\rm cl}\simeq 0.1$--0.2 at $\Sigma_{\rm cl} \sim 100$ \msunpcsq\ to $\epsilon_{*,\rm cl} = 0.93$ in the highest surface density cloud to have been simulated at $\Sigma_{\rm cl} \sim 10^5$ \msunpcsq\ (see Fig.~\ref{fig:estar_tcloud}; left panel). These studies also found that the lifetimes of the clouds varied dramatically with the initial cloud surface density, with the lower surface density ($\Sigma_{\rm cl} \sim 100$ \msunpcsq) clouds being dispersed by feedback from massive stars after only about a free fall time, but denser clouds surviving for up to $\sim 5$--10 free fall times. The longer lifetimes for denser clouds are actually a significant factor in achieving the higher cloud-scale integrated SFE, as the SFE per free fall time (not shown here) has a weaker dependence on the initial cloud density.   Although the cloud lifetimes are longer for the higher density clouds in terms of free fall times, in terms of absolute timescales, the lifetimes of these clouds are shorter due to the strong scaling of free fall time with cloud density. Fig.~\ref{fig:estar_tcloud} shows the cloud lifetimes (defined as the time when 95\% of the final stellar mass has been formed) in the cloud-scale simulations as a function of cloud surface density, in units of the cloud free fall time (middle panel) and in Myr (right panel). 

It has been pointed out in many previous works \citep[e.g.][and references therein]{Grudic2018,Grudic2020,Chevance2023} that the SFE can be qualitatively described remarkably well by a simple analytic model. The basis of the model is the idea that the cloud is dispersed, and star formation halted, when the specific momentum deposition rate (from all sources) exceeds a critical surface density. The critical surface density can be written
\begin{equation}
    \Sigma_{\rm crit} = \frac{\langle \dot{p}/{m_*} \rangle}{\pi G} 
\end{equation}
where $\dot{p}/{m_*}$ is the specific rate of momentum deposition from stellar populations, which here is assumed to include mechanical energy from stellar winds and radiation pressure (but not supernovae). Based on the Starburst99 \citep{Leitherer1999} models\footnote{Other stellar population synthesis models such as BPASS would give similar estimates within 10-20\%.} with a \citet{KroupaIMF} IMF, we estimate $\dot{p}/{m_*} \simeq 30$ km/s/Myr for a zero-age, solar metallicity stellar population (note that this value is roughly constant for the first $\simeq 3$ Myr, until the onset of the first supernova explosions; see e.g. \citet{Hopkins2023} Fig.~2 and \citet{Lancaster2021a} Fig.~3). This yields $\Sigma_{\rm crit}=2176.0$ \msun pc$^{-2}$.

The star formation efficiency over the lifetime of the cloud is then
\begin{equation}
\epsilon_{*, \rm cl} = \frac{\Sigma_{\rm cl}/\Sigma_{\rm crit}}{(1.0+\Sigma_{\rm cl}/\Sigma_{\rm crit})}
\label{eqn:estarcloud}
\end{equation}
The prediction of this simple model for the cloud scale SFE $\epsilon_{*, \rm cl}$ is shown compared with the simulation results in Fig.~\ref{fig:estar_tcloud} (left panel). The model converges to the known values of SFE of a few percent at surface densities typical of local GMCs, and increases to values of close to unity at the highest surface densities considered, in qualitative agreement with the cloud-scale simulation results. Please see Section~\ref{sec:discussion:cloudscale} for a discussion of the reasons that different cloud-scale simulations differ somewhat in their predictions for both the SFE and the cloud lifetimes. 

The cloud free fall time $t_{\rm ff}$ is given by
\begin{equation}
t_{\rm ff} = \left(\frac{3\pi}{32 G \bar{\rho}}\right)^{1/2} = \left(\frac{\pi^2 R_{\rm cl}^3}{8 G M_{\rm cl}}\right)^{1/2}
\end{equation}
where $M_{\rm cl}$ and $R_{\rm cl}$ are the mass and radius of an individual cloud (assumed here to be the same for all clouds; one can think of this as the average cloud mass and radius). 

We fit an empirical function to the cloud lifetimes in the simulations of M24:
\begin{equation}
t_{\rm cl}  = (\tau_{\rm cl,0}\, t_{\rm ff}) \left[(\Sigma_{\rm cl}/\Sigma_{\rm cl, 0})^{-\gamma_1} + (\Sigma_{\rm cl}/\Sigma_{\rm cl, 0})^{\gamma_2}\right]^{-1}
\end{equation}
where we find that $\tau_{\rm cl, 0}=8.64$, $\Sigma_{\rm cl,0}=10^5$ \msun pc$^{-2}$, $\gamma_1=0.4$ and $\gamma_2=0.14$ gives a good fit to the M24 simulation results. We note here that this picture provides a more nuanced interpretation of the interplay between SFE and cloud lifetime, seen in the simulations. The final SFE is determined by the point where feedback can start to overcome gravity. This is higher for higher surface densities, which means that for fixed SFE per free-fall time, it will take more free-fall times to reach this critical point. Thus longer lifetimes are a natural outcome of the finish line being further for the higher-density clouds.

The average gas surface density of the ISM in the galaxy is given by
\begin{equation}
\Sigma_{\rm gas} = \frac{m_{\rm gas}}{\pi r_{\rm disk}^2}
\end{equation}
where $m_{\rm gas}$ is the total mass of gas in the ISM of the galaxy and the characteristic radius of the gas $r_{\rm disk}$ is estimated as described in Section~\ref{sec:scsam}. 

We now assume that the surface density of individual clouds scales with the average ISM surface density, introducing the clumping factor $c$, which we treat as a free parameter:
\begin{equation}
    \Sigma_{\rm cl} = c \Sigma_{\rm gas}
    \label{eqn:sigmacloud}
\end{equation}
The mass of gas in dense, star forming clouds is given by
\begin{equation}
m_{\rm dense} = f_{\rm dense} m_{\rm gas} = N_{\rm cl} M_{\rm cl}
\end{equation}
where $f_{\rm dense}$ is the fraction of the ISM in dense clouds and $N_{\rm cl}$ is the average number of clouds in the galaxy. 

The star formation rate (SFR) per cloud is then given by
\begin{equation}
   \langle SFR \rangle_{\rm cloud} = \frac{\epsilon_{*, \rm cl} M_{\rm cl}}{t_{\rm cl}} 
\end{equation}
and the SFR in the entire galaxy will be
\begin{equation}
    SFR_{\rm gal} = \langle SFR \rangle_{\rm cloud}\, N_{\rm cl} = \langle SFR \rangle_{\rm cloud}\, f_{\rm dense}\, \frac{m_{\rm gas}}{M_{\rm cl}}
\end{equation}
We adopt fiducial values of $c=1$ and $M_{\rm cl}=10^5$ \msun\ throughout this work. The cloud radius can then be computed based on the cloud surface density, given by Eqn.~\ref{eqn:sigmacloud}.

A brief aside on the notional parameter $f_{\rm dense}$: physically, it is intended to represent the fraction of the ISM that is in bound proto-clusters or in dense star forming clouds. It may be related to the H$_2$ fraction in the galaxy, but the relationship between star formation and H$_2$ at very high redshifts is uncertain. Moreover, simulation-based recipes for H$_2$ fraction \citep[e.g.][]{Gnedin2011,Polzin2024} have not been calibrated at the high gas surface densities relevant to this work. The fraction of gas or star formation occurring in bound clusters may also be related to the onset of internal gravitational instability within the galaxy (i.e. the Toomre Q parameter), which requires information about the structural and kinematic properties of galaxies that are not really available in our SAM. We searched the literature for numerical simulations that could give us insights into this quantity and its dependence on galaxy properties that are predicted by our SAM, but could not find any galaxy-scale simulations with sufficiently high resolution that probe the extremely high gas surface densities relevant to this work. Since we do not know how to estimate $f_{\rm dense}$ based on galaxy properties that are available in our SAM, we have chosen to adopt a fixed value for this parameter in all galaxies at all cosmic times, and to vary this value in different model runs to explore the sensitivity of the model predictions to this parameter. However, in reality, of course we expect that $f_{\rm dense}$ will vary from galaxy to galaxy (or even within a galaxy), and may have trends with galaxy properties, which could introduce an effective redshift dependence. 

We now implement the new star formation model in the SAM in the following manner. We divide the intervals between snapshots in our merger trees into timesteps of $\Delta t \sim 3$ Myr. In each timestep, a new population of clouds is assumed to begin to form stars with the efficiency described above. Any gas that is remaining at the end of the timestep (including gas that is in clouds) can be ejected by supernova driven winds, according to the standard prescription described in Section~\ref{sec:scsam}, where the SFR used to drive the wind is taken from the \emph{previous} timestep, reflecting the fact that there is a delay of $\sim 3$ Myr between the onset of star formation and the first SN explosions. Mass conservation is enforced at every timestep. 

\begin{table}
\begin{center}
\centering
\begin{tabular}{lccc}
  \hline\hline
  name & SF recipe & $f_{\rm dense}$ & winds \\
  \hline
Y25 & GKBig2 (S15) & N/A & on \\
KS & Kennicutt-Schmidt (S08) & N/A & on \\
KS-nowind & Kennicutt-Schmidt (S08) & N/A & off \\
\hline
cloud-fd=0.1 & cloud based (\S 2.3) & 0.1 & on \\
cloud-fd=0.5 & cloud based (\S 2.3) & 0.5 & on \\
cloud-fd=1 &  cloud based (\S 2.3) & 1.0 & on \\
cloud-fd=1-nowind &  cloud based (\S 2.3) & 1.0 & off \\
\hline
\end{tabular}
\caption{Summary of the model variants presented in this paper. }
\label{tab:models}
\end{center}
\end{table}

\subsection{Dust Model}
\label{sec:dust}
For illustrative purposes, we adopt a very simple approach for modeling attenuation by dust. We assume that the dust to metal ratio is $\zeta_{\rm dust}=0.08$ \citep{Behrens2018,Pallottini2022}, so that $m_{\rm dust} = \zeta_{\rm dust} Z_{\rm cold}$, where $Z_{\rm cold}$ is the metallicity of the cold ISM gas in absolute units (here we assume solar metallicity is 0.015), which is calculated by the SC SAM using a simple instantaneous recycling approximation treatment of chemical evolution \citep{Somerville2008,Somerville2015}. We then estimate the optical depth of the dust as 
\begin{equation}
\tau_{\rm dust} = \kappa_{\rm V} \, \frac{m_{\rm dust}}{\pi\, r_{\rm dust}^2}
\end{equation}
where we assume that the radial extent of the dust is the same as that of the star forming gas in the ISM, $r_{\rm dust} = r_{\rm disk}$. We adopt $\kappa_{\rm V}=8.55 \times 10^3$ cm$^2$ g$^{-1}$, which is typical for dust with composition similar to the Milky Way \citep{Draine2001}. We then compute the attenuation in the rest-UV, in magnitudes, according to $A_{\rm UV} = (A_{\rm UV}/A_V) 0.921 \tau_{\rm dust}$, which is applicable for dust that is distributed in a screen around the sources. We refer to this as our ``screen dust'' model.

\citet[][hereafter F24]{Ferrara2024} have suggested that when the bolometric luminosity of the galaxy exceeds an effective Eddington luminosity, radiation pressure on the coupled dust-gas medium could eject the dust from the galaxy or push it out to larger distances such that the optical depth drops to a negligible value \citep[see also][]{Ziparo2022,Ferrara2023,Fiore2023}. F24 then show that, under the assumption that the galactic potential is dominated by the stars, and a fixed conversion factor between SFR, UV luminosity L$_{\rm UV}$, and bolometric luminosity L$_{\rm bol}$, the Eddington condition translates to a critical specific SFR (sSFR$_{\rm crit}$) above which galaxies may become ``attenuation free".  We investigate a similar idea applied in post-processing to our SAM results, where we adopt the screen model as described above for galaxies with sSFR$<$sSFR$_{\rm crit}$, and set $A_{\rm UV}=0$ in galaxies with sSFR$>$sSFR$_{\rm crit}$. We refer to this as our ``screen+sSFR$_{\rm crit}$" dust model. 

We adopt $(A_{\rm UV}/A_V) = 2.5$, typical of a Calzetti-type attenuation law \citep{Calzetti2000}.  We note that recent studies have shown that some high-redshift galaxies have attenuation curves that are well fit by the Calzetti functional form \citep{Markov2025,Fisher2025}. However, it is also known that there is a very broad dispersion in attenuation curve shape even in nearby galaxies \citep{Salim2020}. Some of this variation arises even for a fixed underlying extinction curve (reflecting a non-evolving dust composition and grain size distribution), due to scattering and variations in the dust column density along the line of sight, as demonstrated by radiative transport modeling of galaxies in cosmological hydrodynamic simulations \citep{Narayanan2018,Sommovigo2025}. Additional variation in the attenuation curve could arise from changes in the dust chemistry or grain size distribution \citep{Hirashita2020,Narayanan2025}. 

\begin{figure*}
    \includegraphics[width=2\columnwidth]{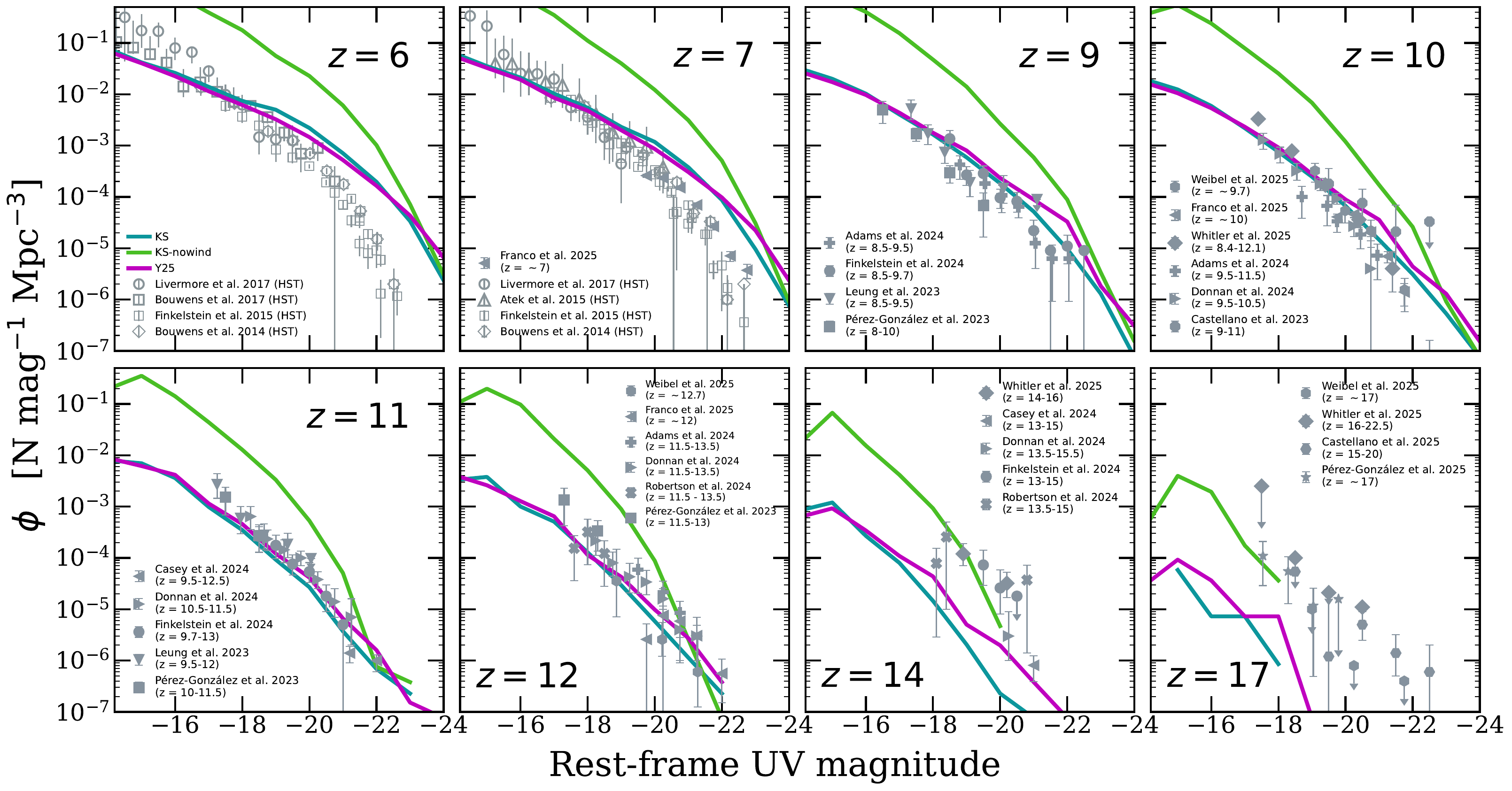}
    \caption{
    \textbf{Baseline models (no dust or enhanced bursts):} Rest-UV luminosity functions at redshift $6 < z < 17$. Symbols show a compilation of observational luminosity function estimates, as specified in the figure legend. Also shown are the predictions from the \textbf{KS}, \textbf{Y25}, and \textbf{KS-nowind} models (without the inclusion of dust or enhanced bursts) as labeled in the figure legend (see text for details). As shown by previous works, the baseline models (\textbf{KS} and \textbf{Y25}) reproduce the observations well from $6 \lesssim z \lesssim 10$, but underproduce the observed UVLF estimates at $z\gtrsim 12$. The excess in the model predictions at bright UV luminosities at $z \sim 6$--7 is plausibly due to dust attenuation, which has not been accounted for in the model predictions shown here. The \textbf{KS-nowind} model, which has supernova driven winds switched off, better reproduces luminous galaxies at $z\gtrsim 12$, but overproduces lower-luminosity galaxies. 
    }
    \label{fig:UVLF_base}
\end{figure*}

\begin{figure*}
    \includegraphics[width=2\columnwidth]{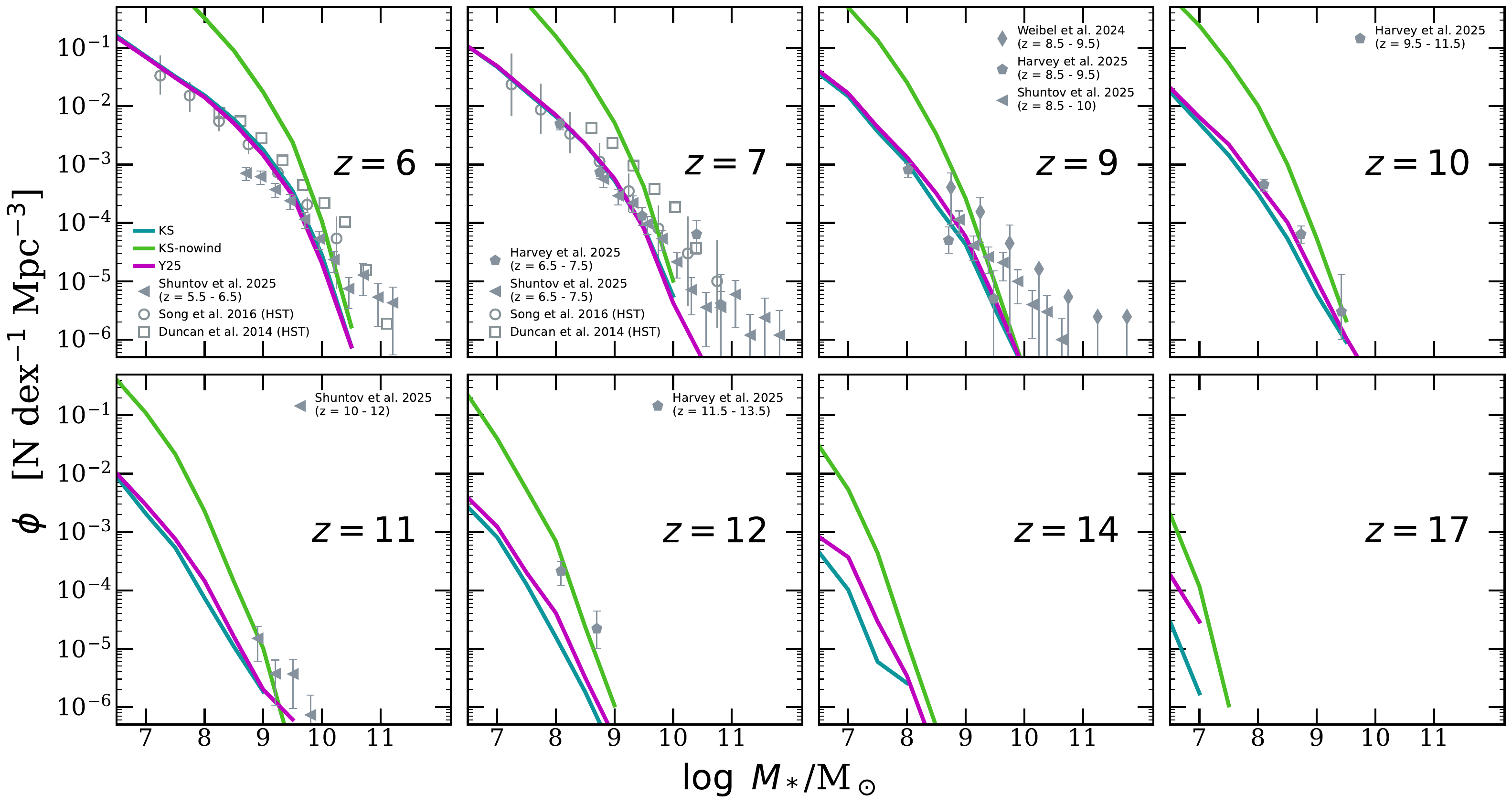}
    \caption{\textbf{Baseline models:} Stellar mass functions (SMF) at redshift $6 < z < 17$. Symbols show a compilation of observational SMF estimates, as specified in the figure legend. Also shown are the predictions from the \textbf{KS}, \textbf{Y25}, and \textbf{KS-nowind} models as labeled in the figure legend (see text for details). The baseline models (\textbf{KS} and \textbf{Y25}) reproduce the observational SMF estimates quite well at $z\sim 6$--7 for lower mass galaxies ($m_* \lesssim 10^9$ \msun), but underproduce more massive galaxies, and also underproduce the observational estimates at $z\gtrsim 10$. The \textbf{KS-nowind} model, which has supernova driven winds switched off, better reproduces massive galaxies at $z\gtrsim 11$, but overproduces lower-mass galaxies. 
    }
    \label{fig:SMF_base}
\end{figure*}

\begin{figure*}
    \includegraphics[width=2\columnwidth]{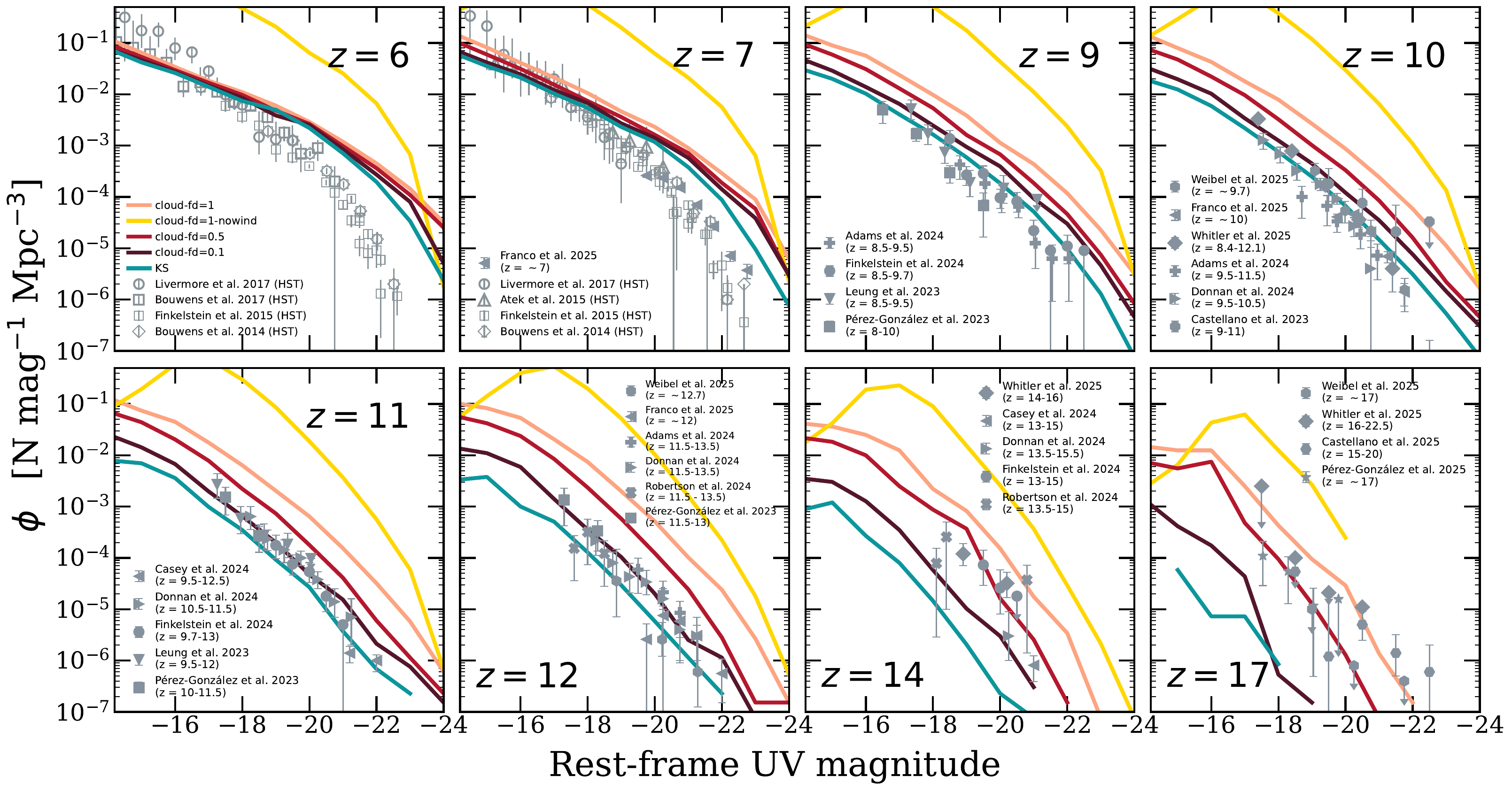}
    \caption{\textbf{DMSFE models (no dust or enhanced bursts):} Rest-UV luminosity functions at redshift $6 < z < 17$. Symbols show a compilation of observational luminosity function estimates, as specified in the figure legend. Also shown are the predictions from the \textbf{KS} model and the density modulated star formation efficiency (DMSFE) models, with different fractions of the ISM in dense star forming clouds ($f_{\rm dense}=0.1$, 0.5, 1.0) as labeled in the figure legend (see text for details). The DMSFE models for all values of $f_{\rm dense}$ give similar predictions to our baseline \textbf{KS} model at $z=6$. The DMSFE models predict higher overall UVLF normalizations relative to the baseline \textbf{KS} model by a factor that increases with redshift, and with the assumed value of $f_{\rm dense}$. The excess in the model predictions at bright UV luminosities at $z \sim 6$--7 is plausibly due to dust attenuation, which has not been accounted for in the model predictions shown here. Assuming that dust attenuation can be neglected at the higher redshifts, we find good agreement with the observations with $f_{\rm dense}=0.1$ at $z\sim 11$--12 and with $f_{\rm dense}=0.5$ at $14<z<17$, suggesting that the observations can easily be accommodated within this family of models with physically reasonable values of $f_{\rm dense}$.  The \textbf{cloud-fd=1-nowind} model has unity cloud-scale star formation efficiency, and has galaxy-scale winds switched off, and likely represents a nearly maximal prediction of the UVLF.  
    }
    \label{fig:UVLF_cloud}
\end{figure*}

\begin{figure*}
    \includegraphics[width=2\columnwidth]{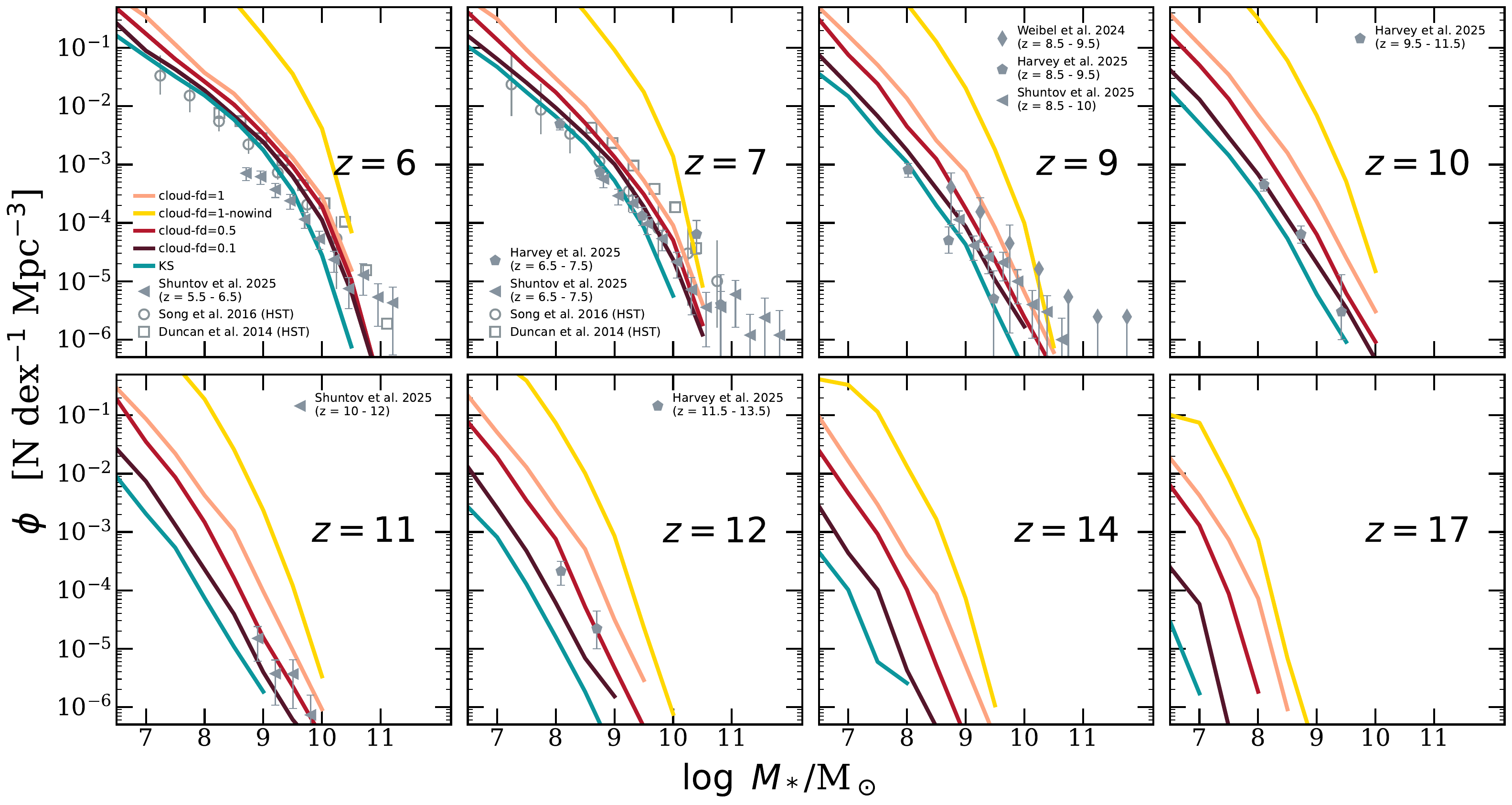}
    \caption{\textbf{DMSFE models:} Stellar mass functions (SMF) at redshift $6 < z < 17$. Symbols show a compilation of observational SMF estimates, as specified in the figure legend. Also shown are the predictions from the \textbf{KS} model and the density modulated star formation efficiency (DMSFE) models, with different fractions of the ISM in dense star forming clouds ($f_{\rm dense}=0.1$, 0.5, 1.0) as labeled in the figure legend (see text for details). We find good agreement with the observations with $f_{\rm dense}=0.1$--0.5 at $z\sim 9$--12. Because they form more stars at early times, these models also produce better agreement with the observational estimates at the massive end of the SMF at $z\sim 6$--7, though they are still lower than the highest nominal observational estimates at very high stellar masses $m_* \gtrsim 10^{10}$ \msun. This could be impacted by the limited number of the rarest, most massive halos in the relatively small volume of our largest N-body simulation. 
    }
    \label{fig:SMF_cloud}
\end{figure*}

\subsection{Modeling bursty star formation in post-processing}
\label{sec:burstmodel}
Although our SAM predicts star formation histories that are somewhat ``bursty'', and includes starbursts triggered by galaxy-galaxy mergers, we may be missing sources of stochasticity connected to small scale processes in the ISM that are not resolved or properly treated by our simple framework. In numerical simulations that partially resolve a multi-phase ISM, star formation can be quite stochastic \citep{Sun2023,Pallottini2023}, operationally leading to a broad distribution of galaxy UV luminosity at a given halo mass, which can be written $P(M_{\rm UV}|M_h)$. This causes galaxies in lower mass halos to be ``boosted'' in luminosity and become observable, where they would have otherwise been too faint to be detected at a given survey sensitivity. This can have a significant impact on the expected UV luminosity functions and $M_{\rm UV}$-limited UV luminosity densities and has been suggested as an explanation for the slower-than-expected decline of the number of UV-luminous galaxies towards higher redshift \citep{Mason2023,Shen2023,Sun2023,Kravtsov2024,Gelli2024}. However, it is unclear that the amount of ``burstyness" predicted by high resolution numerical simulations such as FIRE or SERRA is sufficient to relieve the observational tension \citep{Pallottini2023,Gelli2024}.

In order to explore this effect within the context of our models, we adopt the approach of \citet[][hereafter G24]{Gelli2024} and implement additional stochasticity in $M_{\rm UV}$ in post-processing, assuming that $P(M_{\rm UV}|M_h)$ has a Gaussian functional form and using the G24 halo mass dependent expression for the dispersion $\sigma_{\rm UV}(M_h)$, where: 
\begin{equation}
\sigma_{\rm UV} = a \log(M_{\rm h}) + b
\end{equation}
with $a=-0.34$ and $b=4.5$. This expression is derived from the FIRE-2 simulation suite, and is assumed to be independent of redshift. We sample this distribution with 200 independent draws for each galaxy to compute the UVLF, and report the median. We note that this approach may in fact \emph{overestimate} the amount of `burstyness', since the SC-SAM star formation histories have some intrinsic burstyness due to feedback and mergers \citep{Iyer2020}. Moreover, we emphasize that integrating bursty star formation self-consistently within our SAM would likely lead to somewhat different results than this post-processing approach, as these bursts are expected to have an impact on other galaxy properties such as gas content and metallicity. 

\section{Results}
\label{sec:results}
In this section, we present predictions from our models for the relationship between galaxy radius and UV magnitude at $z=9$--13 (\S\ref{sec:sizes}), and for distribution functions of key galaxy properties at $z=9$--17, including rest-UV luminosity functions and stellar mass functions, in our baseline models and our new density modulated SFE model suite (\S\ref{sec:DMSFEimpact}). To construct all of these quantities, we combine results from our four \gureft\ volumes with the larger VSMDPL volume (details of the procedure are described in Yung et al. in prep).  In \S\ref{sec:sfe}, we present the instantaneous and global SF efficiencies as a function of halo mass and redshift, and in \S\ref{sec:dustimpact} and \S\ref{sec:burstimpact} we show the impact of adding dust and bursty star formation in post-processing. We show the combined effects of density modulated SFE, dust, and bursts in \S\ref{sec:alltogethernow}. We provide a summary of the different models shown throughout this section in Table~\ref{tab:models}.
These include \textbf{Y25}: a model with identical physics to the one published in Y24, but with an updated and more accurate computation of the photometry, as described in Yung et al. (2025); \textbf{KS}: the baseline SC-SAM model, but using the single gas phase Kennicutt-Schmidt model for star formation instead of the H$_2$-based star formation model that was used in Y24 and \textbf{Y25}; \textbf{KS-nowind}: the \textbf{KS} model, but with the supernova driven galaxy-scale winds switched off completely; \textbf{cloud-fd=0.1}, \textbf{cloud-fd=0.5}, \textbf{cloud-fd=1.0}: the new density modulated SFE (DMSFE) model with the dense gas fraction $f_{\rm dense}=0.1$, 0.5, and 1.0 respectively; \textbf{cloud-fd=1.0-nowind}: the DMSFE model with $f_{\rm dense}=1.0$ and galactic winds switched off. 

\subsection{Galaxy sizes}
\label{sec:sizes}
 Observed galaxy sizes in the rest-UV provide a constraint on the surface density of star-forming gas in these galaxies, and therefore we consider it important that our model adopts galaxy sizes that are in reasonable agreement with the available observations. Although galaxies at $z\gtrsim 9$ are very compact, the majority of them are resolved in the rest-UV with multiple pixels by JWST. Sizes are reported for the first epoch of the CEERS sample in \citet{Finkelstein2023}, for the COSMOS-Web sample in \citet{Casey2024}, and for four galaxies in the JADES sample in \citet{Robertson2024}. In addition, a study of galaxy sizes for nine public extragalactic JWST fields taken in Cycle 1 was presented in \citet{Morishita2024}. There seems to be a systematic difference between the sizes taken from the CEERS, COSMOS-Web and JADES papers and the size-$M_{\rm UV}$ relation presented by \citet{Morishita2024}, the origin of which is currently unclear to us (see Fig.~\ref{fig:sizemag}). 
 
We note that there are factor of a few uncertainties in converting from the stellar 3D half-mass radius (provided by our SAM) and the projected UV half-light radius, plotted for the observations. \citet{Morishita2024} show that the optical half-light radius is the same as the UV half-light radius for these high redshift galaxies, which is in contrast to the situation for lower redshift galaxies, but is not too surprising given the young stellar populations in these objects. The conversion from 2D projected light to 3D light is given by \citep{Behroozi2022}:
 \begin{equation}
 r_{\rm proj} = \frac{r_{\rm 3D}}{f_{\rm 3D} f_{\rm b/a}}
 \end{equation}
 where $f_{\rm 3D}=1$ for disks, and $f_{\rm b/a}$ depends on the aspect ratio (ratio of minor to major axis). For randomly oriented disks, the average value of $(b/a)=0.5$, yielding $f_{\rm b/a} = 0.75$ \citep{Behroozi2022}, and thus $r_{\rm proj} \simeq 1.33 r_{\rm 3D}$. However, the 3D shapes of galaxies at these redshifts are unknown, although there is evidence that galaxies at intermediate redshift do \emph{not} show axis ratio distributions that are consistent with randomly oriented disks or spheroids \citep{Pandya2024}. 
 
 Bearing in mind these uncertainties, we make the simple assumption that the disk scale radius\footnote{Note that the half-mass radius plotted is $r_{1/2} = 1.68 r_{\rm disk}$.} $r_{\rm disk} = f_r \lambda_h R_{\rm vir}$, with $\lambda_h$ the dimensionless spin parameter of the halo as measured by the {\sc rockstar} halo finder (\citealp{Peebles1969} definition). Adopting $f_r=2$, we find reasonable agreement with the observed size-magnitude relation for $z\sim 7$--13 galaxies from the CEERS, COSMOS-Web, and JADES compilation.  Given that the most probable value of the halo spin parameter at $z\sim 10$ is $\lambda_h \simeq 0.03$ \citep{Yung2024a}, this implies an average ratio of the galaxy radius to the halo virial radius of $r_{\rm disk}/R_{\rm h} \sim 0.06$. This is about a factor of two to three larger than the values inferred from abundance matching at lower redshifts of $z\sim 3$--7 \citep{Shibuya2015,Somerville2018}. Fig.~\ref{fig:sizemag} shows the intrinsic rest UV magnitude (without a correction for dust attenuation) versus the galaxy stellar 3D half-mass radius for the \textbf{KS} model, compared with the rest-UV magnitude and projected UV half-light radii from the observational samples described above. 
 
If we assume that the radius of the gas in the galaxy is $r_{\rm gas} = 0.06 R_{\rm h}$ at all times, and the cold ISM gas mass is $m_{\rm gas} = f_{\rm gas} f_b M_{\rm h}$  then we can estimate the gas surface density $\Sigma_{\rm gas}$ for a halo of any given mass as a function of redshift.  Fig.~\ref{fig:gasdensityz} shows $\Sigma_{\rm gas}$ in this simple toy model as a function of redshift for a halo with $M_{\rm h} = 10^{11}$ \msun\ and different assumed values of $f_{\rm gas}=0.2$, 0.5, and 1.0; we note that in this simple picture, $\Sigma_{\rm gas} \propto M_{\rm h}^{1/2}$. This toy model motivates the qualitative expectation that 1) ISM gas may be up to two orders of magnitude higher in surface density at ultra-high redshift than in the local Universe and 2) galaxies may be expected to cross into the regime where $\Sigma_{\rm gas}$ exceeds the critical value of a few $\times 10^3$ \msunpcsq, where stellar feedback becomes unable to effectively regulate star formation, at around $z\sim 10$ \citep[see also][]{Dekel2023}.

Fig.~\ref{fig:Sigma_Gas} shows the gas surface density $\Sigma_{\rm gas}$ in our baseline \textbf{KS} semi-analytic model. Here, the cold ISM gas mass is computed by the model based on the inflow rate of fresh gas into the ISM, the rate of gas consumption by stars, and the rate at which the ISM is ejected by supernovae-driven winds, as described in \S\ref{sec:scsam}. Therefore, the gas density is a more complex function of halo mass and redshift, leading to somewhat less dramatic evolution in $\Sigma_{\rm gas}$ than that predicted by the toy model with fixed $f_{\rm gas}$. 
 
\subsection{Impact of the density modulated star formation efficiency}
\label{sec:DMSFEimpact}
Figure~\ref{fig:UVLF_base} shows the intrinsic rest-UV luminosity function without dust attenuation at different redshifts from $z=9$--17 for the ``baseline'' SC-SAM models (\textbf{Y25} and \textbf{KS}) compared with a set of observational estimates. At $z\sim 6$--7 we show UVLF estimates from pre-JWST observations \citep{Bouwens2015,Finkelstein2015,Bouwens2017,Livermore2017,Atek2015}. At $z \gtrsim 8$, we show select recent UVLF estimates from JWST observations \citep{Leung2023,Perez-Gonzalez2023,Adams2024,Casey2024,Donnan2024,Finkelstein2024,Robertson2024,Whitler2025,Perez-Gonzalez2025,Castellano2025, Franco2025, Weibel2025}. 

We first note that although the \textbf{Y25} model shown here contains almost exactly the same underlying physics as the published results of Y24, it incorporates a more accurate treatment of the stellar population synthesis and photometry than the UVLF shown in Y24, and produces better agreement with observations out to $z\sim 11$--12 (more details will be provided in Yung et al., in prep).  We also note that the \textbf{KS} model produces slightly but systematically fewer very UV-luminous galaxies than the \textbf{Y25} model. This is due to the fact that the slope of the Kennicutt-like relation between gas surface density and SFR surface density steepens above a critical surface density in the \textbf{Y25} model, while the \textbf{KS} model assumes a constant slope of $N_K = 1.5$. As discussed and shown in \citet{Somerville2015} and \citet{Yung2019a}, this results in shorter depletion times in early galaxies, which tend to have gas surface densities higher than the critical value. Moving to higher redshift, the baseline models predict fewer galaxies than some works have reported, by up to a full two orders of magnitude by $z=14$.  We note that especially at the highest redshifts shown in our comparison ($z\sim 14$--17), our predictions at the brightest magnitudes can be impacted by the limited volume of our largest N-body simulation (VSMDPL), as we are missing extremely rare, massive halos. 

Lastly, we see that if we completely switch off supernova driven galaxy scale outflows (\textbf{KS-nowind}), the model produces more UV-luminous galaxies, in excess of the observations at $z=9$--12, but still potentially slightly undershooting the observations for bright galaxies at $z\gtrsim 14$, depending on which of the bright-end $z\sim 14$ UVLF estimates one believes. Moreover, switching off the SN-driven winds tends to lead to a much steeper UVLF slope than that seen in the observations, and a general excess of  
galaxies (except at the brightest magnitudes) at $z\lesssim 12$. This suggests that simply making supernova driven winds weaker will not fully solve the problem. 

In Fig.~\ref{fig:SMF_base} we show the stellar mass function for the same set of models and same redshifts as in Fig.~\ref{fig:UVLF_base}. The takeaway messages are similar; the baseline models (\textbf{Y25} and \textbf{KS}) produce similar results, with \textbf{Y25} predicting slightly more galaxies with large stellar masses. The baseline model with galactic winds switched off (\textbf{KS-nowind}) predicts steeper stellar mass functions. Fig.~\ref{fig:SMF_base} also shows observational estimates of the stellar mass function at $z\sim 9$--12. While these estimates are quite uncertain, it is notable that the baseline models seem to produce fewer massive galaxies than the observational estimates taken at face value, particularly at $z\gtrsim 10$.

\begin{figure*}
    \includegraphics[width=2\columnwidth]{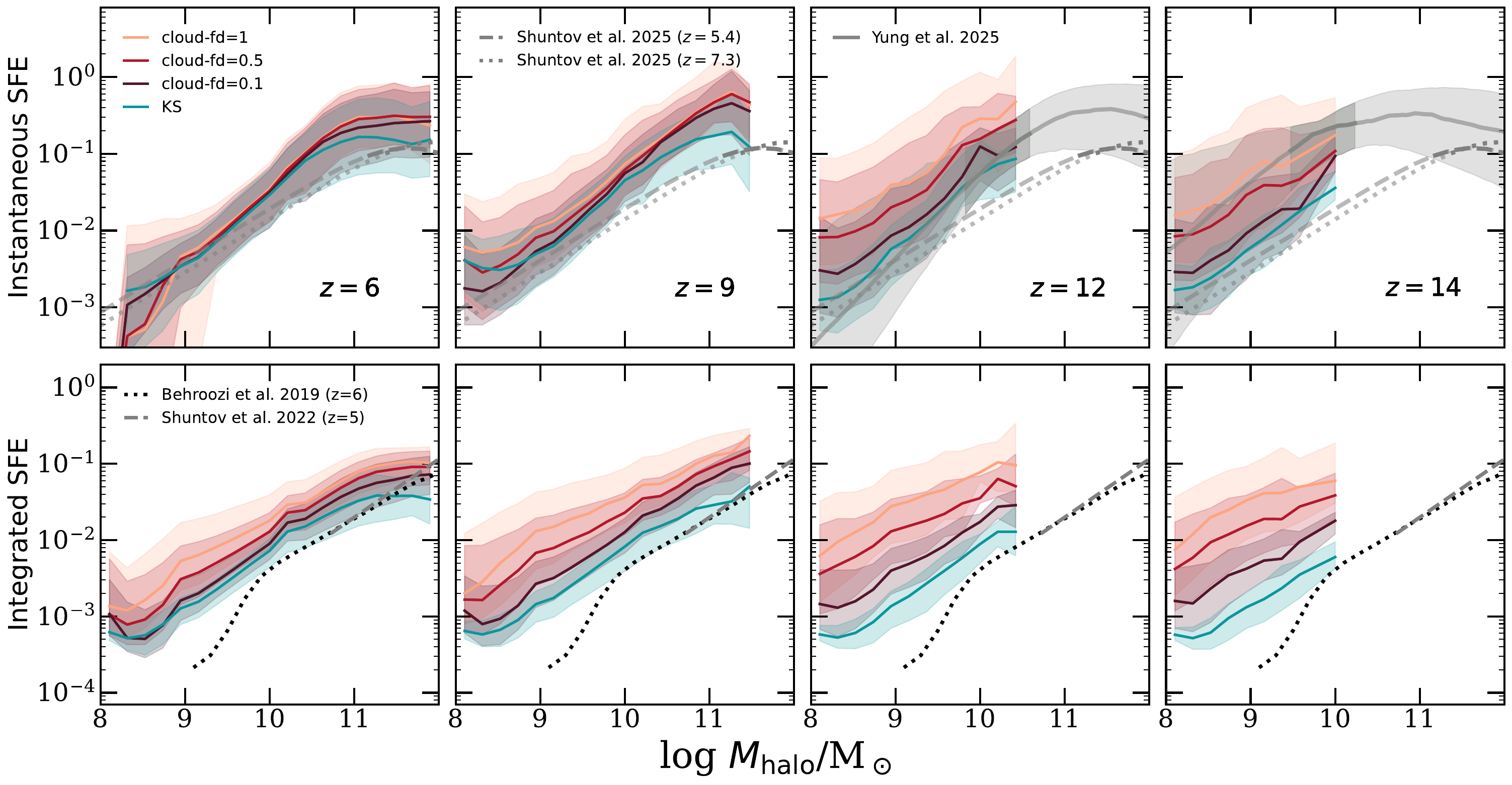}
    \caption{Instantaneous and integrated star formation efficiency (SFE; see text for definitions) as a function of halo mass for redshifts $z=6$, 9, 12 and 14. We show the \textbf{KS} model and the DMSFE models with three values of the dense gas fraction $f_{\rm dense}=0.1$, 0.5, and 1.0 as shown on the figure legend. Solid lines show the medians and the shaded regions show the dispersion in values over the predicted galaxy population (16th and 84th percentiles). In addition, we show empirical estimates of these quantities from abundance matching models (gray and black lines), where the darker lines indicate where there are existing constraints from observed UVLF or SMF. The dotted and dashed lines are the $z\sim 5$--7 empirical model results shown repeated in each panel to guide the eye. Here, the shaded regions represent the 16th and 84th percentiles of the posterior. This figure illustrates that when comparing implied SFE at different redshifts, it is important to compare at a roughly common halo mass, since the empirical global SFE is a strong function of halo mass, and the halos that host observable galaxies shift towards lower masses at high redshift.   
    }
    \label{fig:SFE}
\end{figure*}

\begin{figure*}
    \includegraphics[width=2\columnwidth]{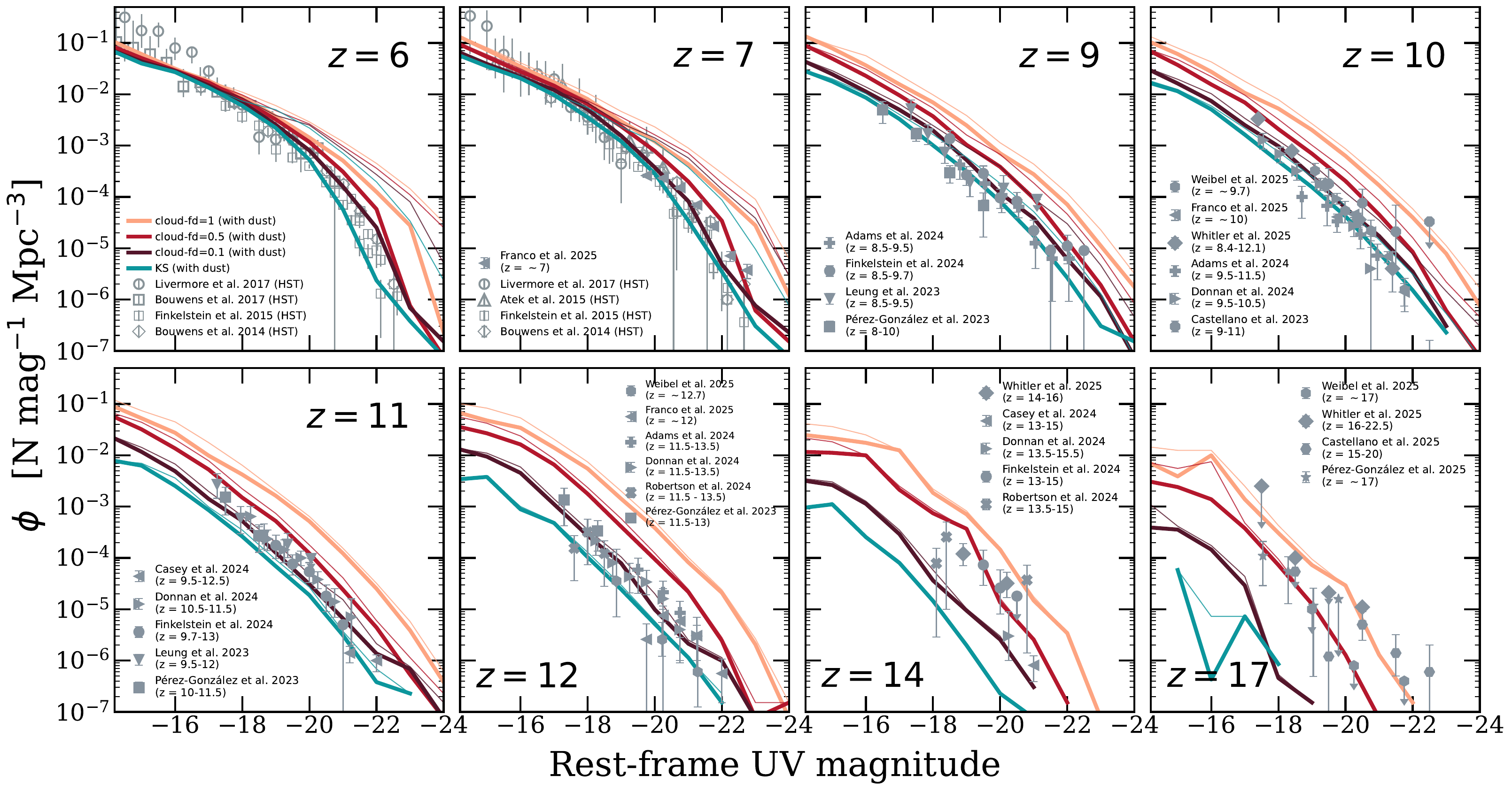}
    \caption{\textbf{DMSFE models with dust:} Rest-UV luminosity functions at redshift $6 < z < 17$. Symbols show a compilation of observational luminosity function estimates, as specified in the figure legend. Also shown are the predictions from the \textbf{KS} model and the density modulated star formation efficiency (DMSFE) models, with different fractions of the ISM in dense star forming clouds ($f_{\rm dense}=0.1$, 0.5, 1.0) as labeled in the figure legend (see text for details). Light solid lines show the models without any correction for dust attenuation, and heavy lines show the UVLF after correcting for dust attenuation using the screen+sSFR$_{\rm crit}$ model (see text). This model predicts a negligible impact of dust attenuation at $z\gtrsim 11$, and an increasing impact at lower redshifts.
    }
    \label{fig:UVLF_dust}
\end{figure*}

\begin{figure}
    \includegraphics[width=\columnwidth]{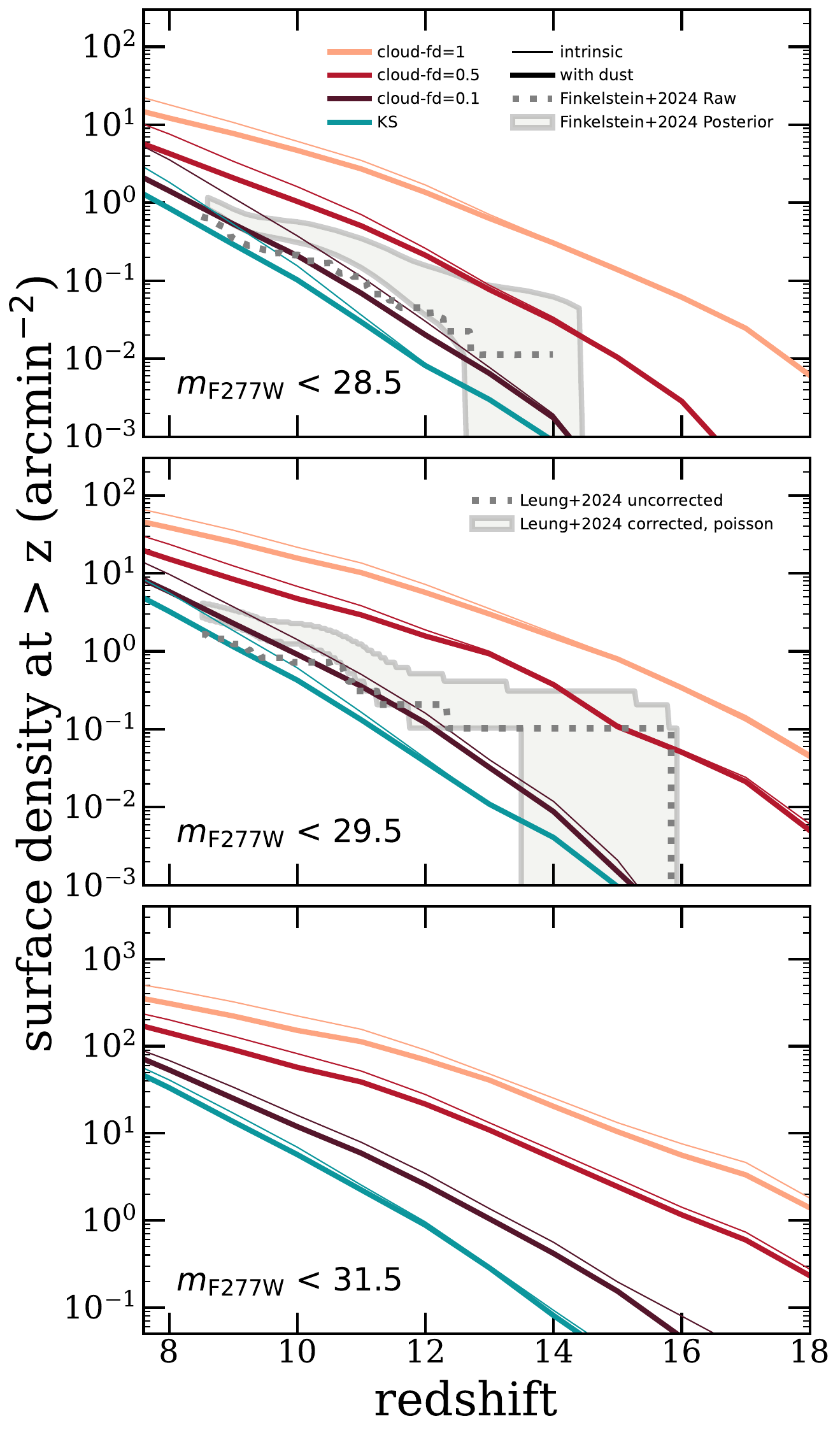}
    \caption{\textbf{DMSFE models with dust:} Cumulative number counts of galaxies per arcmin$^2$ on the sky with redshifts above the one plotted on the x-axis and apparent magnitude brighter than $m_{\rm 277W}<28.5$ (top), $m_{\rm 277W}<29.5$ (middle), and $m_{\rm 277W}<31.5$ (bottom). The dotted grey lines show the uncorrected number counts from the CEERS (top; \citealp{Finkelstein2024}) and NGDEEP (middle; \citealp{Bagley2023,Leung2023}) surveys and the shaded grey regions show the results from these surveys after a correction for incompleteness; the bottom panel represents a possible future ultra-deep JWST survey. 
    We show the predictions from the \textbf{KS} model and the density modulated star formation efficiency (DMSFE) models, with different fractions of the ISM in dense star forming clouds ($f_{\rm dense}=0.1$, 0.5, 1.0) as labeled in the figure legend (see text for details).  Lighter lines show the intrinsic counts and the darker lines show the counts including dust attenuation using the screen+sSFR$_{\rm crit}$ model.      
    }
    \label{fig:counts_dust}
\end{figure}

\begin{figure*}
    \includegraphics[width=2\columnwidth]{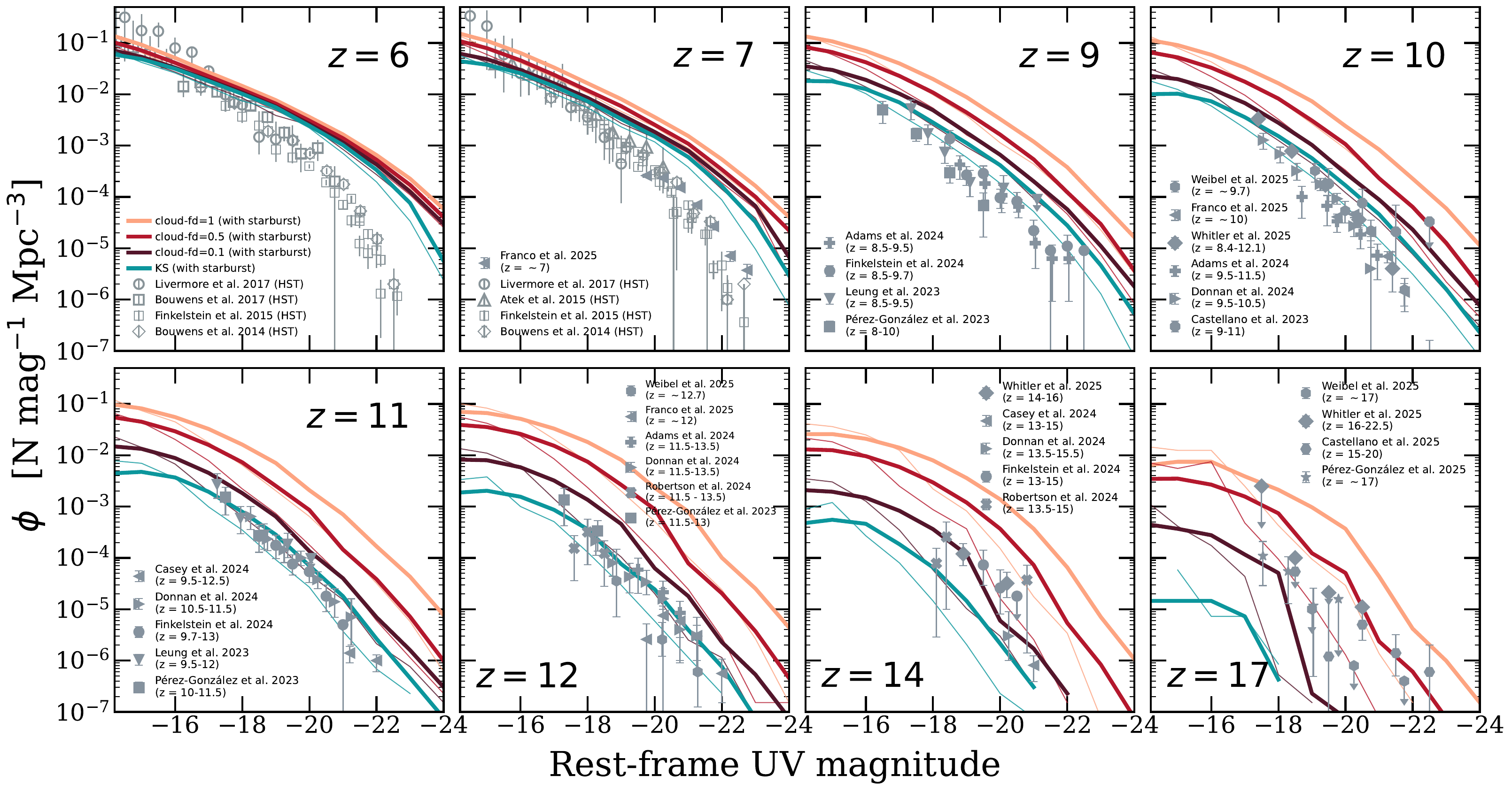}
    \caption{\textbf{DMSFE models with enhanced bursts (no dust):} Rest-UV luminosity functions at redshift $6 < z < 17$. Symbols show a compilation of observational luminosity function estimates, as specified in the figure legend. Also shown are the predictions from the \textbf{KS} model and the density modulated star formation efficiency (DMSFE) models, with different fractions of the ISM in dense star forming clouds ($f_{\rm dense}=0.1$, 0.5, 1.0) as labeled in the figure legend (see text for details), all without dust attenuation modeling included. Light solid lines show the models without enhanced bursty star formation; heavier lines show the impact of including enhanced bursts in post-processing (see text for details). Enhanced bursts preferentially boost the bright end of the UVLF at higher redshifts.
    }
    \label{fig:UVLF_burst}
\end{figure*}

Fig.~\ref{fig:UVLF_cloud} shows the predictions of our new density modulated star formation efficiency (DMSFE) model for the UVLF without dust attenuation and without additional bursts added in post-processing. Different colored lines show different values of $f_{\rm dense}$, the fraction of the total ISM in dense clouds or proto-star-clusters. The \textbf{cloud-fd=0.1}, \textbf{cloud-fd=0.5}, and \textbf{cloud-fd=1} models adopt galactic wind feedback parameters that are the same as in the baseline/published model, while the \textbf{cloud-fd=1-nowind} model has galactic scale winds switched off. The \textbf{cloud-fd=0.1} model is consistent with the observations from $z\sim 9$--12 but underpredicts at least some of the observational estimates at $z\sim 14$, while the \textbf{cloud-fd=0.5} model is consistent with or higher than the observations out to $z \sim 17$. The \textbf{cloud-fd=1} and  \textbf{cloud-fd=1-nowind} models are considerably higher than all of the observational estimates (except for the brightest galaxy candidates at $z\sim 17$ reported by \citet{Castellano2025}, with which they are consistent). No single value of $f_{\rm dense}$ seems to be able to match the observations at all redshifts. However, in reality we expect this quantity to depend on galaxy properties such as density, metallicity, and internal structure, which could introduce an effective dependence on cosmic time (see \S~\ref{sec:discussion} for more discussion). 
 
Fig.~\ref{fig:SMF_cloud} shows the stellar mass function for the DMSFE models for the same set of values of $f_{\rm dense}$. As expected, since star formation is more efficient at early times in the DMSFE models with higher values of $f_{\rm dense}$, larger amounts of stellar mass are in place at $z\sim 6$--12, where there are stellar mass estimates. Once again, the \textbf{cloud-fd=0.1} and \textbf{cloud-fd=0.5} models are in the same range as the observational estimates at $z\sim 9$--12. None of our models reproduce the highest estimated number densities of very massive galaxies $m_* \gtrsim 10^{10}$ \msun at $z\sim 6$--9, but these are based on a small number of objects and our results could be impacted by the relatively small volume of our largest N-body simulation volume.

\subsection{Integrated and instantaneous efficiencies in the DMSFE model}
\label{sec:sfe}
There are multiple different definitions of star formation efficiency in the literature. A commonly used one is the cloud scale SFE per free fall time, $\epsilon_{\rm ff}$, which is defined as the fraction of gas on some scale (e.g. in a star forming cloud, local patch of ISM, or a whole galaxy) that is turned into stars per free fall time. This is often used as an input into numerical simulations of galaxy formation. In this section, we examine the emergent \emph{global} star formation efficiencies in our model galaxies in different models. This can also be defined in two ways: 1) the \emph{instantaneous} SFE$_{\rm inst} \equiv \dot{m}_* /(f_b \dot{M}_{\rm h})$, where $\dot{m}_*$ is the SFR, $f_b$ is the universal baryon fraction, and $\dot{M}_{\rm h}$ is the total mass accretion rate into the halo.  2) the \emph{integrated} SFE$_{\rm int} \equiv m_* /(f_b M_{\rm h})$ where $m_*$ is the stellar mass and $M_h$ is the host halo mass. If the instantaneous SFE were constant in time, then the instantaneous and integrated SFE would be the same modulo a factor for recycled stellar ejecta from massive stars. However, in our models the instantaneous SFE can vary with time due to the dependence on galaxy properties that change with time. 

Fig.~\ref{fig:SFE} shows the instantaneous SFE in the top row, and the integrated SFE in the bottom row. To compute the former, we use the instantaneous SFR from the SAM (i.e., averaged over 3 Myr). We show empirical models based on abundance matching from \citet{Shuntov2025} and \citet{Yung2025} in the top row and from \citet{Behroozi2019} and \citet{Shuntov2022} in the bottom row, compared with the DMSFE models with $f_{\rm dense}=0.1$, 0.5, and 1.0. Interestingly, although the surface densities at $z\gtrsim 10$ lead to very high cloud scale efficiencies $\epsilon_{*, \rm cl} \gtrsim 0.5$, the effective galaxy scale efficiencies are much smaller, due to the galaxy scale supernova driven winds, which remove gas from the ISM so that it is not available for star formation. We also notice that the \textbf{KS} model and the models with different values of $f_{\rm dense}$ yield almost the same global instantaneous SFE at $z=6$, especially in halos with masses $M_{\rm h} \lesssim 10^{11}$ \msun. This is because when the galaxy scale gas depletion time $t_{\rm dep}$ (defined as the time it would take for all of the gas in the ISM to be consumed, at the present SFR, $t_{\rm dep} \equiv m_{\rm gas}/\dot{m}_*$) is short compared with the age of the Universe, the galaxy scale SFE is mainly set by the mass loading of the galactic winds (which in our SAM scales with the rotation velocity of the galaxy). As $t_{\rm dep}$ becomes comparable to or longer than the age of the Universe, the amount of gas that can be converted into stars becomes increasingly limited by the (cloud scale) SFE \citep[see the discussion and references in][]{Somerville2015}. 

Fig.~\ref{fig:SFE} also helps illustrate why the DMSFE model does not reproduce the UV LF \emph{evolution} for any fixed value of $f_{\rm dense}$. Although the DMSFE models do predict stronger evolution in the SFE than the baseline models, it is not as strong as what empirical models \citep{Shuntov2025,Yung2025} indicate would be required to account for the observed UVLF evolution in the absence of other effects, such as a changing stellar light-to-mass ratio. The recent empirical study of \citet{Yung2025} suggests that the instantaneous SFE at $M_{\rm h} \sim 10^{10}$ \msun\ must have decreased by a factor of $\sim 4$ from $z\sim 14$--12, which is a rather staggeringly rapid decrease considering that this represents only 70 Myr of cosmic time. This perhaps hints at the need for other physical processes to explain the observed behavior, which we discuss and explore later. 

\subsection{Impact of dust on the UV luminosity function evolution}
\label{sec:dustimpact}
In this subsection we investigate the impact of including a correction for dust attenuation on our UVLF predictions. Fig.~\ref{fig:UVLF_no_thres} shows our \textbf{cloud-fd=0.1} DMSFE model with the screen dust model and the screen+sSFR$_{\rm crit}$ model, in which we assume that the value of the UV attenuation $A_{\rm UV}=0$ whenever the specific star formation rate (sSFR) exceeds a given critical value sSFR$_{\rm crit}$. For the results shown we adopt sSFR$_{\rm crit}=25$ Gyr$^{-1}$ as proposed by \citet{Ferrara2024}. The standard screen dust model predicts rather high values of $A_{\rm UV}$ even at redshifts $z\gtrsim 10$, where observed UV spectral slopes suggest that there is very little reddening \citep{Cullen2023,Morales2024}. Interestingly, the screen+sSFR$_{\rm crit}$ model brings our model UVLF into fairly good agreement with the observed ones at $z\sim 6$--9, and predicts a decreasing impact of dust attenuation as redshift increases, where by $z\sim 12$ the impact of dust is negligible. This is because, in addition to the overall decreasing dust abundance, the distribution of sSFR in our SAMs increases back to earlier cosmic times, due to the increased accretion rates onto halos and into galaxies, leading to a larger fraction of galaxies being cleared of dust. 

Fig.~\ref{fig:UVLF_dust} shows the screen+sSFR$_{\rm crit}$ model applied to our baseline \textbf{KS} model and to our family of DMSFE models with different values of $f_{\rm dense}$. As already seen, the \textbf{cloud-fd=0.1} DMSFE model agrees well with the observations at $6 \lesssim z \lesssim 12$, but the \textbf{KS} model with this dust model applied slightly overpredicts UV-bright galaxies at $z\lesssim 9$ and underpredicts them at $z\gtrsim 12$. The other models (\textbf{cloud-fd=0.5} and \textbf{cloud-fd=1}) overpredict the UVLF over the range $6 \lesssim z \lesssim 12$ under the assumptions of this simple dust treatment. 

Fig.~\ref{fig:counts_dust} shows the cumulative number counts of galaxies per square arcmin on the sky with redshifts greater than the value plotted on the x-axis, for three different apparent magnitude limits. The top panel is for m${\rm F277W}<28.5$, which is typical of medium deep surveys such as CEERS \citep{Finkelstein2025}; the middle panel shows m${\rm F277W}<29.5$ which is representative of deeper surveys such as NGDEEP and JADES, and the bottom panel shows m${\rm F277W}<31.5$ which could potentially be reached by a future ultra-deep JWST survey. Here, we show the baseline \textbf{KS} model and the cloud models with $f_{\rm dense}=0.1$, 0.5, and 1.0. For each model, we show the predictions without dust and with the screen+sSFR$_{\rm crit}$ model as before. The screen+sSFR$_{\rm crit}$ dust model clearly makes the slope of the decline of the number of luminous galaxies shallower between $z\sim 6$--12, but still does not bring either our baseline model (which is already too low at $z\gtrsim 10$ even without dust) nor any of the cloud models into perfect agreement with the  observations. The \textbf{cloud-fd=0.1} and \textbf{cloud-fd=0.5} models with the screen+sSFR$_{\rm crit}$ dust model do roughly bracket the observed counts in this redshift range for both magnitude limits, however, which is promising. 

\subsection{Impact of bursty star formation on the UV luminosity function evolution}
\label{sec:burstimpact}
Fig.~\ref{fig:UVLF_burst} shows the impact of adding bursts in post-processing as described in Section~\ref{sec:burstmodel} to our baseline \textbf{KS} model and the three DMSFE model variants with different values of $f_{\rm dense}$. We find that adding bursts brings our \textbf{KS} model into agreement with the observed UVLF at $z\sim 11$--12, but this level of ``burstyness'' alone is not enough to bring this model into agreement with the observed UVLF at $z \gtrsim 14$. 
However, adding bursts to our \textbf{cloud-fd=0.1} model brings this model into plausible agreement with the bright-end UVLF at $z\sim 14$ -- but at the expense of making it a bit high at lower redshifts $z \lesssim 12$. This burst prescription does preferentially boost the number of bright galaxies at higher redshifts, because of the assumed broadening of $P(M_{\rm UV}|M_h)$ towards lower halo masses, and the shift in the most common halos towards lower masses at higher redshift due to the evolving halo mass function. However, the effect as currently implemented is not strong enough alone to bring any of our models into agreement with the nominal observed shallow decline of UV-bright galaxies over $14 \lesssim z \lesssim 6$.

\subsection{Putting it all together: evolution of UV luminous galaxies and the cosmic SFR density}
\label{sec:alltogethernow}
\begin{figure*}
    \includegraphics[width=2\columnwidth]{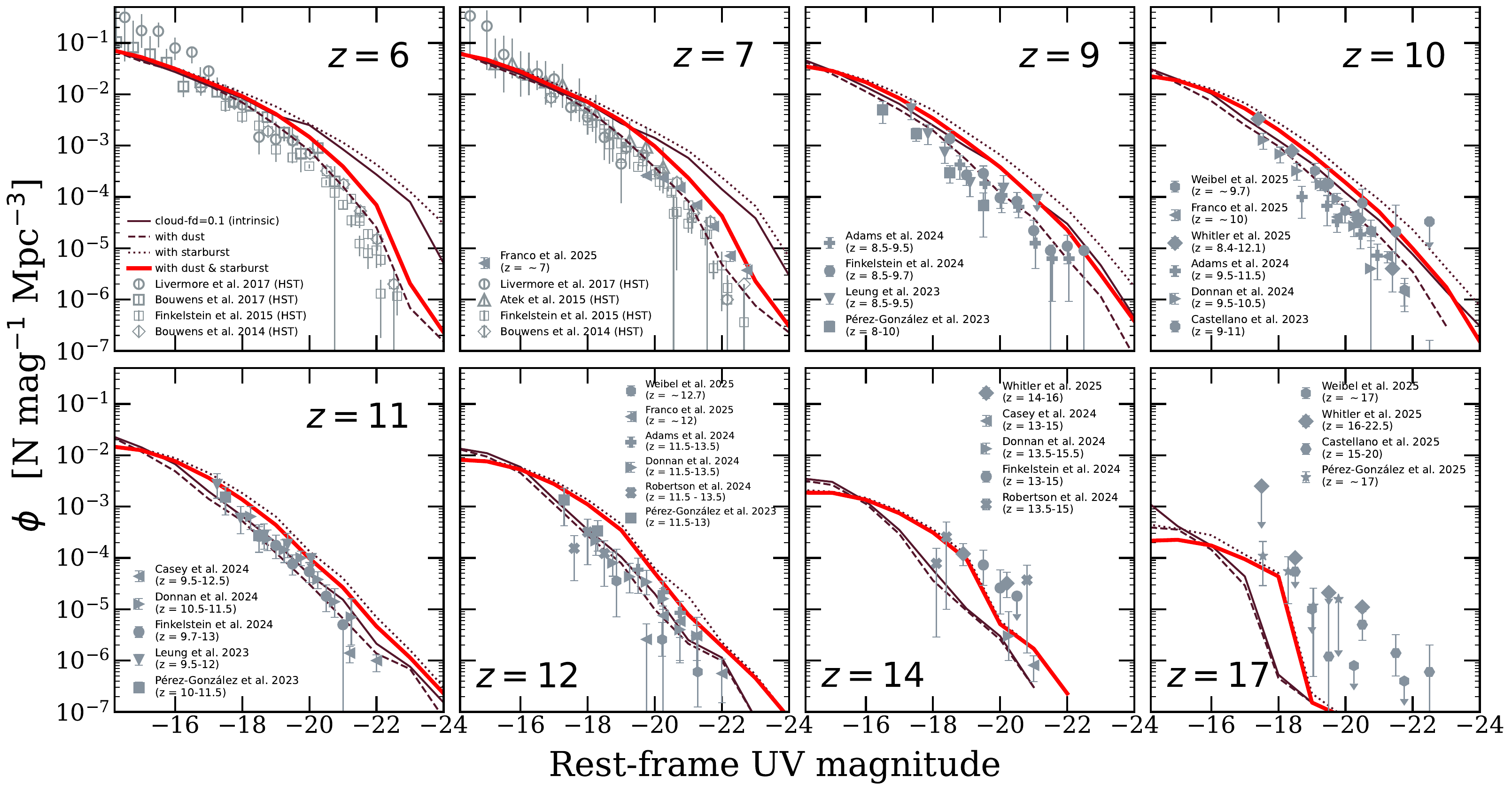}
    \caption{\textbf{DMSFE model with dust and bursts:} Rest-UV luminosity functions at redshift $6 < z < 17$. Symbols show a compilation of observational luminosity function estimates, as specified in the figure legend. Also shown are the predictions from the density modulated star formation efficiency (DMSFE) model with $f_{\rm dense}=0.1$. Lighter brown lines show the models without dust and without bursts. Dashed lines show the models with the inclusion of dust, but no bursts. Dotted lines show the impact of including bursts, but not dust. Red solid lines show the impact of including both bursts and dust. 
    }
    \label{fig:UVLF_dust_n_burst}
\end{figure*}

\begin{figure}
    \includegraphics[width=\columnwidth]{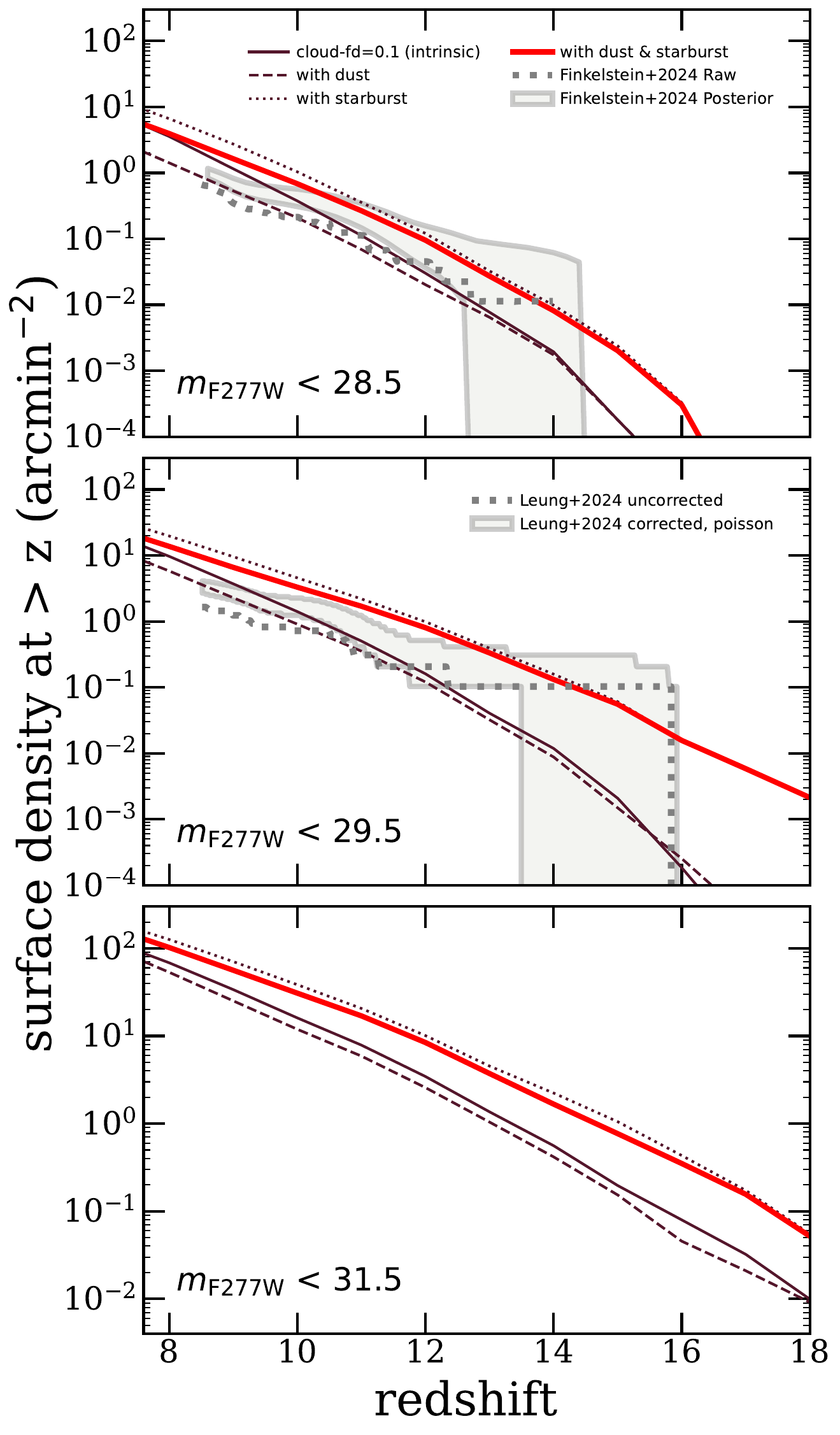}
    \caption{\textbf{DMSFE model with dust and bursts:} Cumulative number counts of galaxies per arcmin$^2$ on the sky with redshifts above the one plotted on the x-axis and apparent magnitude brighter than $m_{\rm 277W}<28.5$ (top), $m_{\rm 277W}<29.5$ (middle), and $m_{\rm 277W}<31.5$ (bottom). The dotted grey lines show the uncorrected number counts from the CEERS (top; \citealp{Finkelstein2024}) and NGDEEP (middle; \citealp{Bagley2023,Leung2023}) surveys and the shaded grey regions show the results from these surveys after a correction for incompleteness; the bottom panel represents a possible future ultra-deep JWST survey. Lighter brown lines show the models without dust and without bursts. Dashed lines show the models with the inclusion of dust, but no bursts. Dotted lines show the impact of including bursts, but not dust. Red solid lines show the impact of including bursts and dust. 
    }
    \label{fig:counts_dust_n_burst}
\end{figure}

To summarize so far: we have shown that the DMSFE model boosts the number density of UV luminous galaxies at $z\gtrsim 9$ by a factor that depends on the assumed value of the fraction of gas in dense SF clouds ($f_{\rm dense}$), but that no single value of $f_{\rm dense}$ is able to match the observed UVLF evolution from $z\sim 14$--6. We also showed that introducing a dust model in which dust is ejected from galaxies with sSFR above a critical value causing them to be ``attenuation free'', as suggested by F24, makes the rise in the number of UV-bright galaxies from $z \sim 12$--6 shallower, in better agreement with observations, but it does not alleviate the tensions at $z \gtrsim 12$ for our baseline \textbf{KS} model, which is already low even without dust. We also showed that introducing an enhancement in bursty star formation in post-processing, using the halo mass dependent burst amplitude dependence suggested by \citet{Gelli2024} based on the FIRE simulations \citep{Sun2023}, preferentially boosts the number density of UV luminous galaxies at high redshift, but again, this affect alone is not enough to bring any of our models into agreement with the observed UVLF evolution. In this section, we show the results of combining our DMSFE model with the screen+sSFR$_{\rm crit}$ dust model and the halo mass dependent post-processed burst model. To our knowledge, this is the first time that all three proposed effects have been simultaneously incorporated into the same modeling framework. 

Fig.~\ref{fig:UVLF_dust_n_burst} shows the UVLF from $6\lesssim z \lesssim 17$ for four permutations of our DMSFE \textbf{cloud-fd=0.1} model: without dust or bursts, with the screen+SFR$_{\rm crit}$ dust model (but no bursts), with the halo mass dependent enhanced burst model (but no dust), and with both the dust and burst model applied. Interestingly, the dust and bursts almost exactly cancel each other out at $z\sim 9$--10, leaving the UVLF normalization a bit too high compared with the observations. As already noted, at $z\gtrsim 11$, nearly all galaxies are predicted to be attenuation free in the screen+SFR$_{\rm crit}$ dust model, so the results revert to the same ones already seen for the \textbf{cloud-fd=0.1}+bursts model --- namely, the model predicts slightly too high a normalization for the UVLF at $z\sim 11$--12, and is in agreement with the lower estimates of the UVLF at $z\sim 14$. 

Fig.~\ref{fig:counts_dust_n_burst} shows the cumulative number density of galaxies brighter than three apparent magnitude limits, similar to Fig.~\ref{fig:counts_dust}. The \textbf{cloud-fd=0.1} model with dust and bursts comes fairly close to reproducing the observed number density of both bright ($m_{\rm F277W}<28.5$) and fainter ($m_{\rm F277W}<29.5$) galaxies compared to the observations from CEERS and NGDEEP, respectively. Interestingly, the problem now is that the models \emph{overproduce} the number of galaxies at $z\sim 9$ by about a factor of two (and also overproduce the counts at lower redshifts, as can be inferred from Fig.~\ref{fig:UVLF_dust_n_burst}). Fig.~\ref{fig:counts_dust_n_burst_KS} shows the same comparison for the \textbf{KS} model, illustrating that this model with the screen+SFR$_{\rm crit}$ dust model plus FIRE-2 calibrated bursts has difficulty matching the observations out to $z\sim 14$.

\begin{figure}
    \includegraphics[width=\columnwidth]{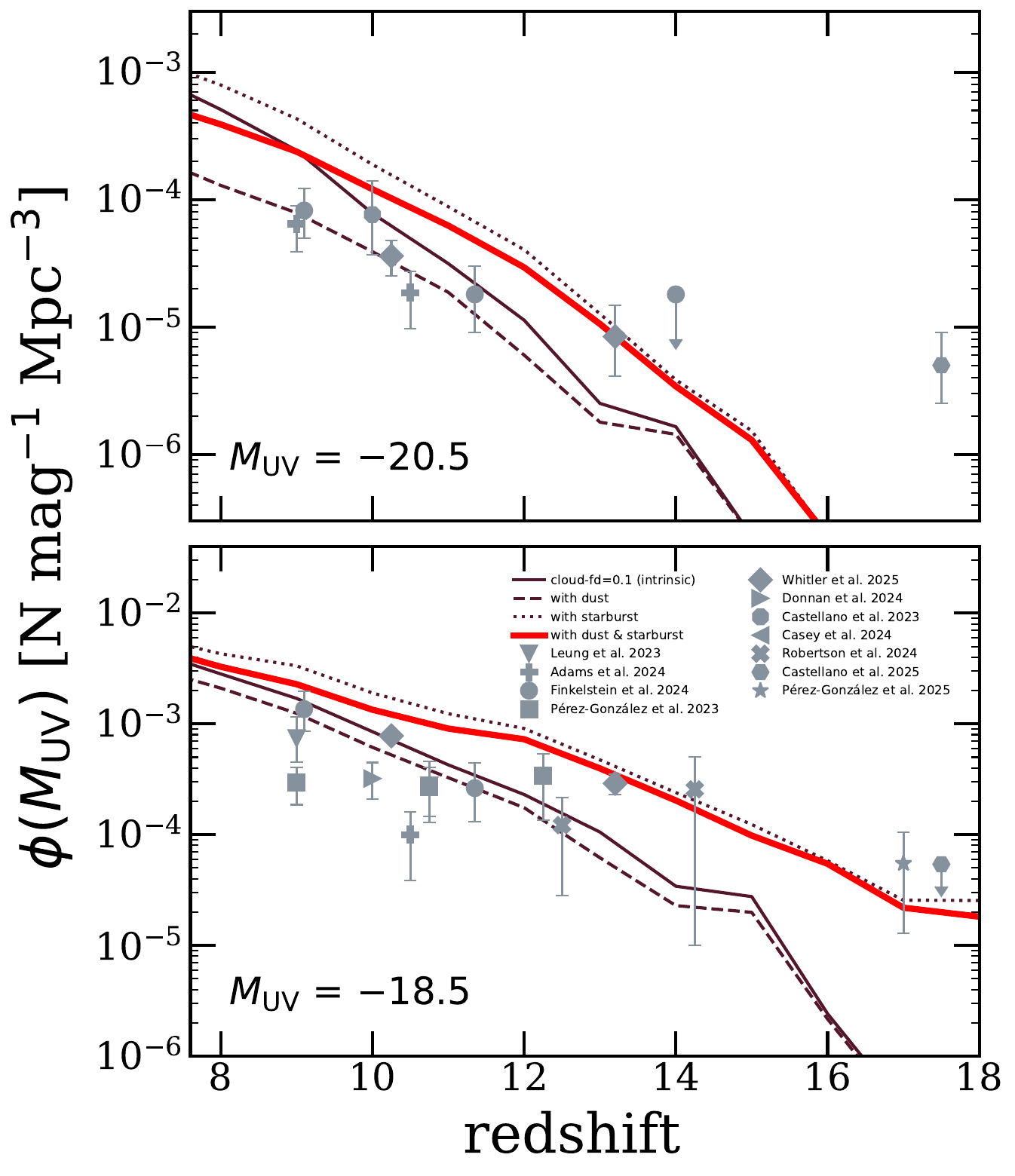}
    \caption{The comoving number density of galaxies with $M_{\rm UV}=-20.5$ (top) and $M_{\rm UV}=-18.5$ (bottom) as a function of redshift. Symbols show observational estimates, and predictions from the DMSFE model with \textbf{cloud-fd=0.1}, with different permutations of dust and bursts included in post-processing are shown with the same line types as in Fig.~\ref{fig:counts_dust_n_burst}. The evolving impact of dust attenuation makes the evolution of $\phi(M_{\rm UV})$ shallower at $z\lesssim 12$ but has little impact at higher redshifts; the preferential boosting of bright galaxies at higher redshifts by enhanced starbursts leads to shallower evolution in $\phi(M_{\rm UV})$ at $z \gtrsim 12$. The combined effects of our DMSFE model, dust, and bursts leads to better agreement with the observed shallow decline in $\phi(M_{\rm UV})$ with increasing redshift. 
    }
    \label{fig:nvsz}
\end{figure}

Fig.~\ref{fig:nvsz} shows a similar comparison, but now for the comoving number density of galaxies $\phi(M_{\rm UV})$ in two different rest-UV absolute magnitude bins, $M_{\rm UV}=-20.5$ and $M_{\rm UV}=-18.5$, as a function of redshift, with the same set of permutations of (no dust, no bursts), (dust no bursts), (bursts no dust), and (bursts+dust) for our \textbf{cloud-fd=0.1} model. The qualitative message is the same as in the discussion of the previous figure, but this figure shows a broader compilation of observational estimates, and highlights the fairly broad range of values for $\phi(M_{\rm UV})$ obtained by different studies, particularly at the highest redshifts. Fig.~\ref{fig:nvsz_KS} shows the same comparison for the baseline \textbf{KS} model. Once again, without the boosted star formation efficiency provided by the DMSFE model, this model has difficulty reproducing the observed number density of UV-bright galaxies out to $z\sim 14$. However, the \textbf{cloud-fd=0.1} model with enhanced bursts over-predicts UV luminous galaxies at $z \lesssim 10$.

\begin{figure}
    \includegraphics[width=\columnwidth]{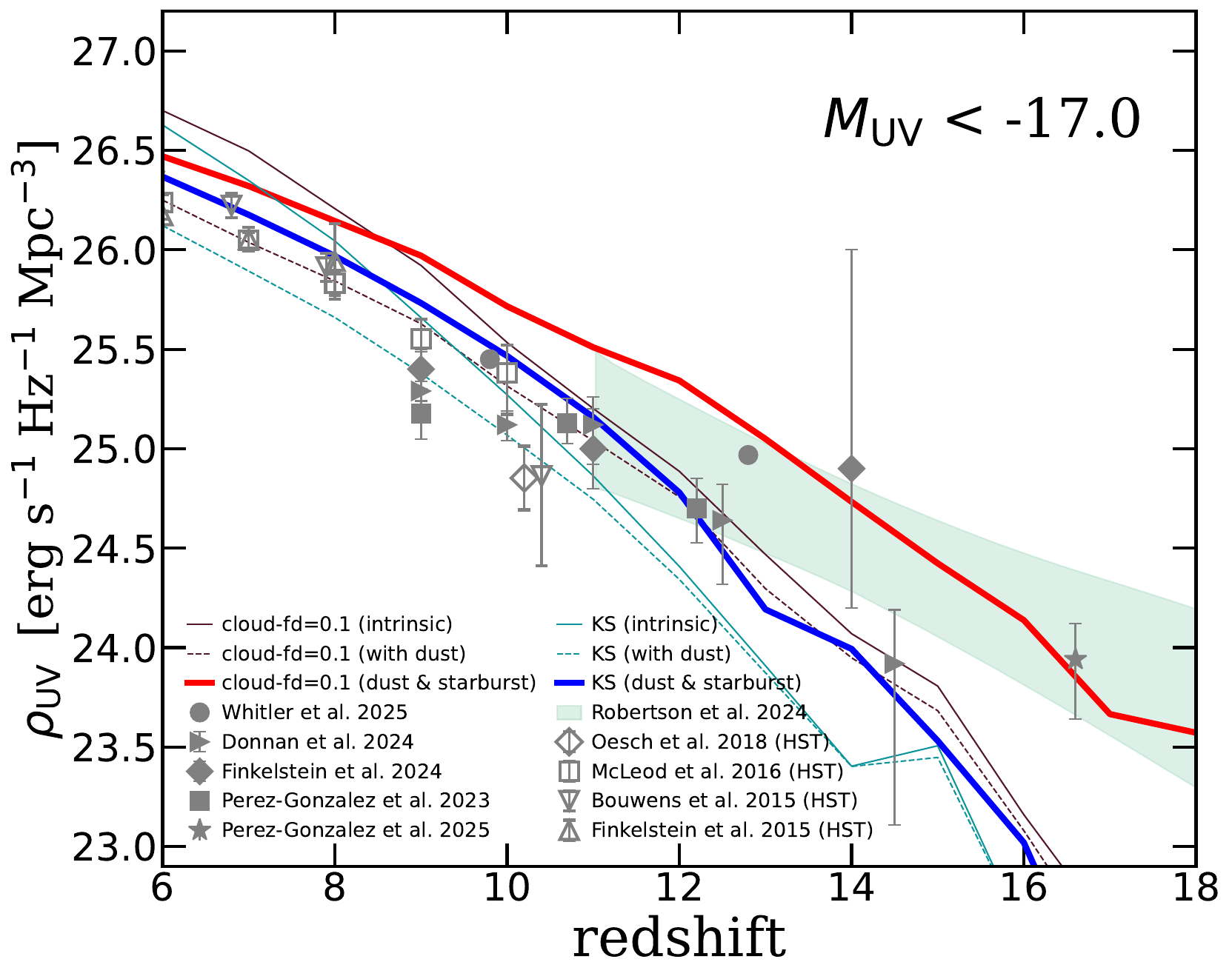}
    \caption{Comoving cosmic UV luminosity density contributed by galaxies brighter than $M_{\rm UV}=-17$, as a function of redshift. We show a compilation of observational estimates from the literature, as specified in the figure legend, along with the \textbf{KS} model (blue \& cyan) and the \textbf{cloud-fd=0.1} model (red \& brown). Both models are shown without dust or bursts (cyan and brown lighter lines) and with both the screen+sSFR$_{\rm crit}$ dust model and the halo mass dependent burst model applied in post-processing (blue and red darker lines). The shallow decline in $\rho_{\rm UV}$ implied by some observational analyses is best reproduced by the \textbf{cloud-fd=0.1}+dust+bursts model.
    }
    \label{fig:rho_UV}
\end{figure}

 Fig.~\ref{fig:rho_UV} shows the comoving cosmic UV luminosity density $\rho_{\rm UV}$ as a function of redshift, integrated above $M_{\rm UV}=-17$. Both the unadorned \textbf{KS} and \textbf{cloud-fd=0.1} models are shown, as well as these models with dust and bursts added in post-processing. Once again, the message is the same: the density modulated star formation efficiency model leads to more star formation in UV-luminous galaxies at early times, but the effect alone is probably not enough to account for the very shallow observed evolution in $\rho_{\rm UV}$. Reduced dust attenuation at higher redshifts, as well as the evolving halo mass function, perhaps leading to larger amplitude bursts of star formation, taken together, may be able to do so, though this is certainly not a unique explanation. 

\section{Discussion}
\label{sec:discussion}

\subsection{Synthesis and interpretation of our results}
To briefly summarize our results, we find that implementing the new DMSFE model, motivated by and calibrated to high resolution cloud-scale simulations, within our cosmological SAM leads to predicted number densities of $z\gtrsim 10$ galaxies that are $\sim 1$ ($f_{\rm dense}=0.1$) to $\sim 2$ ($f_{\rm dense}=1.0$) orders of magnitude higher than our baseline \textbf{KS} model (which assumes a cloud-scale SFE typical of that in nearby galaxies) at $z\sim 12$--14. Moreover, models with physically plausible values of $f_{\rm dense}=0.5$--1 actually predict \emph{higher number densities of UV-luminous galaxies than observational estimates find} at $z\sim 10$--12. Thus the new puzzle seems to be not why we observe so many UV-luminous high redshift galaxies, but why we don't see more of them! 

Although our model does not reproduce the observed time \emph{evolution} in the number density of UV-luminous galaxies with any fixed value of $f_{\rm dense}$, we do not regard this either as surprising nor as a failure of the model, as $f_{\rm dense}$ is expected to depend on galaxy properties and may well have an effective evolution over cosmic time. However, although we tried various plausible scalings of $f_{\rm dense}$ with galaxy properties such as H$_2$ fraction computed using the simulation-calibrated recipes from \citet{Gnedin2011} or \citet{Polzin2024}, or direct scalings with combinations of gas surface density, metallicity, and simple estimates of Toomre Q, we were not able to find a physically motivated scaling of $f_{\rm dense}$ that matched the observed evolution in $\phi(M_{\rm UV}=-20.5)$ from $z\sim 12$--14, taking the current observational estimates at face value. The reason for this is easily understood by returning to Fig.~\ref{fig:Sigma_Gas} --- the gas surface density in our model galaxies simply does not change much over this time interval. The same is true for other physical properties such as metallicities.  Our work suggests that building a solid theoretical understanding of the galaxy-scale conditions that govern the value of $f_{\rm dense}$ is critical for future work. 

We investigated how the role of evolving dust attenuation impacts the UVLF evolution in the context of our models, using a model in which galaxies are assumed to be ''attenuation free'' when their sSFR exceeds a critical value, as suggested by \citet{Ferrara2024} and related works. Since the distribution of sSFR evolves towards higher values at earlier cosmic times in our models, a larger and larger fraction of galaxies are assumed to be attenuation free at higher redshifts, and by $z\gtrsim 12$, dust is predicted to have a negligible impact on the emergent UVLF, with the exception of very faint galaxies, which may have sSFR below the critical value for blow-out, so retain some dust. This has the impact of creating shallower evolution in $\phi(M_{\rm UV})$ and $\rho_{\rm UV}$ from $z\sim 12$--6, bringing \emph{both} our baseline \textbf{KS} and our DMSFE \textbf{cloud-fd=0.1} models into agreement with the observations over this redshift range. This confirms the suggestion of \citet{Ferrara2024} that evolving dust attenuation could contribute to shallower evolution over this redshift interval, within the context of a rather different (and more detailed) modeling framework, which is a non-trivial result. However, our baseline \textbf{KS} model underpredicts the $z\gtrsim 12$ observations even without dust, so this is not a complete explanation for the existing observations, unless the currently reported values of observed galaxy number densities at higher redshift turn out to be significantly over-estimated. 

We investigated one possible mechanism that could contribute to making the evolution of $\phi(M_{\rm UV}=-20.5)$ shallower from $z\sim 12$--14 --- stochastic starbursts with an amplitude that increases with decreasing halo mass. Because the most common halos shift towards lower masses over this time interval due to the rapidly evolving halo mass function, this leads to more low mass halos being ``boosted'' in luminosity at the earlier epochs, and hence higher detected numbers of galaxies. 

Thus, we are suggesting a multi-faceted picture in which 1) density modulation leads to increasing SFE 2) dust attenuation has a decreasing impact on the emergent UV and 3) the amplitude of stochastic bursty star formation increases as a function of increasing redshift. Importantly, we have implemented all of these processes \emph{not} as ad hoc functions of redshift, but based on physical properties (density, accretion rate, halo mass) whose evolution is governed at a fundamental level by cosmological structure formation. While this multi-faceted picture may seem unsatisfying from an Occam's Razor perspective, it is important to consider that there are reasons to expect all of these processes to be physically connected and therefore to occur in tandem. For example, the DMSFE picture, with its higher star formation efficiencies, is expected to lead to strong bursts of star formation, which will cause correspondingly strong bouts of stellar and supernova feedback, leading to rapid modulations in the SFR \citep[see also][]{Dekel2023,Li2024}. 

\subsection{Other observational support for and implications of the DMSFE picture}
Our model is motivated by theory and by results from numerical simulations of individual GMCs or proto-star-clusters. However, there is also observational support for the picture we have presented. Superstar clusters in the nearby Universe, which have surface densities $\Sigma > 10^3$ \msunpcsq, have been observed to have star formation efficiencies of up to 75\% \citep{Turner2017,LindaSmith2020,Costa2021,Villas2020}. Observational evidence for higher SFE on larger scales has also been seen in other systems with high gas surface density, such as nuclear starbursts and merging luminous IR-bright galaxies \citep{Leroy2018,Emig2020,Sun2024}. These conditions are rare in the low redshift universe, but intriguingly, several studies of highly magnified high to ultra-high redshift ($6\lesssim z \lesssim 10$) galaxies behind lensing clusters have shown that these systems are composed of multiple dense, star forming clumps, which have been speculated to perhaps be globular clusters in formation \citep{Vanzella2023,Adamo2024,Mowla2024,Fujimotograpes2024}. These star forming clumps have sizes of less than a parsec, and plausibly have comparable masses ($\sim10^4$--$10^5$ \msun) and surface densities ($\sim 10^4$--$10^5$ \msunpcsq) to the ``clouds'' in our model. \citet{Fujimotograpes2024} estimate that around 50\% of the star formation in the ``Cosmic Grapes'' lensed galaxy at $z\sim 6$ is occurring in the clumps/proto-star-clusters. 

This picture, in which a significant fraction of the star formation in the early Universe took place in (potentially) short-lived, gravitationally bound star clusters, has other interesting implications. For example, JWST spectroscopy has revealed that some (though not all) high redshift galaxies ($6 \lesssim z \lesssim 8$) have enhanced Nitrogen-to-Oxygen (N/O) and Carbon-to-Oxygen (C/O) ratios relative to local H$_{\rm II}$ regions with comparably low Oxygen abundance (O/H) \citep{Topping2024,Marques-Chaves2024}. Similar abundance patterns are, however, seen in globular cluster stars, further suggesting a connection between regions of intense star formation at high redshift and proto-star-clusters \citep{Belokurov2023}. 

These dense star clusters may also represent ideal sites for the formation of Intermediate Mass Black Holes (IMBH), via runaway nuclear core collapse \citep{Inayoshi2020,Marques-Chaves2024,Rantala2024}. High redshift galaxies may contain up to several hundreds of these star clusters, each hosting a BH of $\sim 10^3$\msun. If the star clusters and their IMBH can merge quickly, this could provide a channel for rapid and relatively ubiquitous formation of the seeds of the black holes observed by JWST at $z\sim 5$--6 \citep{Dekel2025}. The implications of rapidly growing BH within high redshift galaxies on the subsequent star formation efficiency as well as on the observable UV luminosity is an important topic for future work. 

\subsection{Caveats and uncertainties of our analysis}
\subsubsection{Observational Uncertainties}
We first discuss the observational uncertainties. The samples of ultra-high redshift galaxies at $z\sim 10$--12 are quite robust. The UVLF has been measured over multiple, relatively large area fields, such that field-to-field variance is not likely to be a dominant source of uncertainty, and a large fraction of the objects have spectroscopic redshifts. However, the UVLF at $z\sim 14$ is much more uncertain, reflected both in the large dispersion between measurements from different groups --- the number density of galaxies at bright magnitudes $M_{\rm UV}\lesssim -20$ from \citet[][PRIMER]{Donnan2024} and \citet[][COSMOS-Web]{Casey2024} is over an order of magnitude smaller than the ones from \citet[][CEERS]{Finkelstein2024} and JADES \citep{Robertson2023,Whitler2025}. This may be in part due to the smaller area of JADES and especially the Jades Origin Field analyzed by \citet{Robertson2023}. In addition, as of this writing, only three galaxy candidates at $z\gtrsim 13$ have been spectroscopically confirmed \citep{Perez-Gonzalez2025,Harikane2025}, and the reliability of photometric redshifts in this epoch is less well understood. Several studies have reported galaxy candidates at even higher redshifts $z\sim15$--30 \citep{Perez-Gonzalez2025,Castellano2025}, but none of these $z\gtrsim 15$ candidates has been spectroscopically confirmed. We have shown the reported UVLF from these studies at $z\sim 17$ in our analysis, but have focussed on the ability of our models to reproduce the observations from $z\sim 6$--14 for this reason.  We do not show predictions at $z\gtrsim 20$, because we do not currently have robust samples of N-body merger trees with the appropriate mass resolution in a large enough volume simulation. We plan to address this issue with a new suite of simulations that is in progress (GUREFT-II; Yung et al. in prep). 

Another observational uncertainty is the impact of dust attenuation on the emergent rest-UV. We know that UV-luminous galaxies at $z\sim 7$ can contain significant amounts of dust, as we see the dust continuum emission with ALMA \citep[e.g.][]{Bouwens2022}. Dust continuum emission from higher redshift galaxies ($z\gtrsim 8$) has not yet been detected (the highest redshift galaxy with a detection of the dust continuum is at $z=8.3$; \citealp{Tamura2019}). Several studies have reported that JWST-selected galaxies show a trend towards bluer UV slopes ($\beta_{\rm UV}$) as redshifts increase towards $z\sim 10$, with the reported slopes at $z\gtrsim 10$ leaving little room for any reddening \citep{Cullen2023,Finkelstein2023,Morales2024}. However, the more precise measurements of $\beta_{\rm UV}$ that are possible with spectroscopy are to date only available for relatively small samples that may be biased towards the UV-brightest objects. It is possible that there is a fainter population of redder objects that is missing in the existing surveys. 

\subsubsection{Modeling the Spectral Energy Distribution}
In this work, we have modeled the spectral energy distribution (SED) using standard techniques, by convolving our predicted star formation and chemical enrichment history with the {\sc BPASS} stellar population models assuming a \citet{Chabrier2003} IMF and the binary star isochrones.  There are multiple significant uncertainties in modeling the emergent light from stellar populations \citep{Conroy2013,Iyer2025}. The isochrones and atmospheres of massive, low metallicity stars are poorly understood, as there are few examples of such stars in the nearby universe \citep{Telford2024}. The fraction of massive stars in multiples is also poorly constrained, but processes associated with binary/multiple star interactions can have a significant impact on the UV \citep{Stanway2018,Gotberg2019}. In this work, we have neglected nebular line and continuum emission from ionized H$_{\rm II}$ regions, but it is known that most high redshift galaxies have very strong nebular line emission that can significantly impact even the broad band NIRCam filter photometry. We plan on including the nebular emission in our models in future work. 

As already mentioned in the introduction, one of the elephants in the room is the stellar initial mass function. The IMF is rather poorly constrained even in the lower redshift Universe. There are multiple physical reasons to expect that the IMF may have been more top heavy (i.e. contain a larger fraction of massive stars per unit stars formed) in early galaxies, including lower metallicities and warmer Cosmic Microwave Background temperatures \citep[see][for a review]{Hennebelle2024}. In fact, the same conditions that we have argued give rise to higher star formation efficiencies due to the weakening impact of massive star feedback could lead to a more top-heavy IMF \citep{Hennebelle2024}. However, in order to explain the observed shallow rise in the number density of UV-bright galaxies and the cosmic SFH, the high-mass slope of the IMF would have to evolve fairly rapidly over a relatively short period of time ($\lesssim 200$ Myr has elapsed between $z\sim 14.5$ and $z\sim 10$). It is unclear whether this is physically reasonable. However, see \citet{Chon2021}, \citet{Chon2022}, \citet{Chon2024} for relevant discussion and modeling results. 

We have also neglected UV light from accreting black holes in our models. \citet{Trinca2024} found that the expected contribution from AGN to the UV in galaxies in the redshift and luminosity range of our study in their CAT semi-analytic models was sub-dominant. Moreover, one might expect that in the dense gas environments that are characteristic of this epoch, much of the UV light from AGN would be obscured. However, as we have just argued above, there might be more efficient seeding and more early BH growth in the DMSFE picture, which would be consistent with the large populations of already fairly massive BH that JWST has discovered at high redshift \citep[e.g.][]{Larson2023,Kocevski2023,Harikane2023,Kocevski2024,Maiolino2024,Matthee2024,Taylor2024,Akins2024}. Therefore this contribution is an open question that is important to model in more detail, which we plan on investigating in future work. 

\subsubsection{Cloud scale physics}
\label{sec:discussion:cloudscale}
The conceptual basis of this work, which is that star formation is more efficient at high gas surface density, is motivated by a simple analytic argument (as summarized in Section \ref{sec:cloudmodel}): as the momentum injection rate from all sources becomes small compared to the attractive forces of gravity, clouds become more difficult to disperse, star forming regions become longer lived in units of the free fall time, and are able to convert a higher fraction of gas into stars \citep[][and references therein]{Grudic2018,Grudic2020}. This is supported by many published numerical simulations across a wide range of scales \citep{Chevance2023}. However, the quantitative values of the cloud-scale SFE as a function of cloud surface density can vary by up to an order of magnitude between different simulations (see Fig.~4 of \citealp{Chevance2023}). This may be partly due to different initial conditions (including different choices of the virialization parameter $\alpha_{\rm vir}$, which describes the balance between kinetic energy and gravitational binding energy, and the initial magnetic field properties), partly due to differences in the numerical treatment of hydrodynamics and radiative transport, and partly due to the combination of physical (feedback) processes that are included (see discussion in \citealp{Chevance2023}, Section 6.1). For example, the simulations of \citet{Lancaster2021} shown in Fig.~\ref{fig:estar_tcloud} include stellar wind feedback but not radiation. It is likely that these simulations produce higher values of $\epsilon_{\rm *}$ at lower cloud surface densities $\Sigma_{\rm cl} \simeq 100$ \msunpcsq\ than observational constraints indicate, and than some other cloud-scale simulations predict, because of their neglect of photoionization, which becomes dominant in this regime \citep{Kimjg2018}. The M24 simulations include UV and IR radiation but not stellar winds or protostellar outflows. Neither of these simulations includes magnetic fields. The cloud scale SFE is also found to have a weak dependence on the virial parameter $\alpha_{\rm vir}$ and the initial cloud mass \citep{Polak2024}. An important physical process that is missing in the cloud scale simulations that we have used to calibrate our model is Lyman-$\alpha$ radiation pressure. This could dominate over all other feedback processes in conditions with low dust abundances, potentially increasing $\Sigma_{\rm crit}$ by as much as an order of magnitude \citep{Smith2017,Tomaselli2021,Nebrin2025}.

Moreover, it is notable that the simulation predictions for $\epsilon_{\rm *}$ are systematically higher at a given $\Sigma_{\rm cl}$ than the prediction of the simple analytic model on which Eqn.~\ref{eqn:estarcloud} is based. This is likely due to the implicit assumption in the simple model of a uniform gas density distribution, whereas in a turbulent medium (as assumed in the cloud-scale simulations), there will be a lognormal distribution of densities, in which most of the mass is in structures with densities higher than $\langle \Sigma_{\rm cl} \rangle \equiv M_{\rm cl}/(\pi R_{\rm cl})$, which will therefore be more difficult to accelerate outwards. More sophisticated analytic models that account for the lognormal distribution of gas densities in GMCs yield higher $\epsilon_*$ values, in better agreement with the numerical simulation results \citep{Thompson2016,Raskutti2016}. Given the uncertainties, we have chosen to retain the simple analytic model, which converges to the observed values of $\epsilon_{*, \rm cl}$ at the lower surface densities characteristic of lower redshift galaxies, and yet still matches the cloud-scale simulations at the higher surface densities.  

Given our discussion above about the possibility that the IMF in these dense, massive clouds may be rich in massive stars, another interesting issue is the impact of the assumed IMF on the cloud scale star formation efficiency. Na\"{i}vely, a more top heavy IMF would have a higher value of $\dot{p}/m_*$, leading to a higher value of $\Sigma_{\rm crit}$, which should yield lower integrated cloud scale star formation efficiencies $\epsilon_{*, \rm cl}$. A top-heavy IMF would also modify the chemical yields from stars, potentially altering dust and cooling physics. \citet{MenonIMF2024} investigated the impact of a top-heavy IMF in an approximate way by boosting the UV luminosity per unit star formation $L_{\rm UV}/m_*$ relative to a Salpeter IMF. In dense clouds with $\Sigma_{\rm cl} \sim 3 \times 10^3$--$3 \times 10^4$ \msunpcsq, they found a relatively weak trend of $\epsilon_{*, \rm cl}$ with the boost in $L_{\rm UV}/m_*$, with a decrease of only about 20\% even for boosts in $L_{\rm UV}/m_*$ of ten times the Salpeter value, and an even smaller effect ($\lesssim 5$ \%) for low metallicity gas ($Z_{\rm gas} =0.01\, $Z$_{\odot}$). This suggests that the expectation of higher SFE in very high density environments may hold even in the presence of a top-heavy IMF.

\subsubsection{Interplay between cloud/cluster, galactic and super-galactic scales}
Perhaps the greatest limitation in modeling high redshift galaxies, and galaxies in general, is the impossibility of explicitly modeling the physics that determines the IMF and GMC-scale star formation efficiency while simultaneously modeling the interplay between GMC scales and their kpc scale environments (nuclear bulges, spiral arms), and the interplay between inflows and outflows on scales of the ISM, Circumgalactic Medium (CGM), and Intergalactic Medium (IGM). For example, the simulations described in Section~\ref{sec:cloudmodel} treat the clouds as isolated entities that are not able to accrete gas from their surroundings. 

Another very large uncertainty is how supernovae regulate galaxy-scale star formation efficiency. The prescription adopted in the SC SAM is extremely simple, with a mass loading factor that is a simple function of the galaxy rotation velocity \citep[see][for details]{Somerville2008}. The parameters of the mass loading function are tuned to match local galaxy properties, but it has been shown that these calibrations are extremely degenerate \citep{Pandya2023}. We do not account for the impact of \emph{energy} deposition by supernovae into the ISM and CGM, which can heat gas, leading to longer cooling times and effective `preventative' feedback \citep{Pandya2023,Carr2023,Voit2024}, and drive turbulence \citep{Pandya2023}. The coupling of SN-ejected energy with the CGM that feeds ultra-high redshift galaxies has not been studied in detail. On the one hand, the metallicity of the CGM is presumably lower, leading to slower radiative losses; on the other hand, if accretion is predominantly via narrow, dense filaments, as suggested by numerical simulations, the coupling may be quite weak. 

\subsection{Comparison with previous work}
In this section, we briefly summarize how our findings compare with those of other recent previous works. We start with the Feedback Free Burst (FFB) model of \citet{Dekel2023}. This model is based on the idea that when the volume density of a halo exceeds a critical value, corresponding to a halo free fall time of less than a few Myr, star formation can occur efficiently, with negligible impact of feedback from massive stars or supernovae. Clearly, the physical ideas behind this model are somewhat similar to those that motivate our DMSFE model, however, the implementation of the ideas into a cosmological framework has some significant differences. For example, \citet{Li2024} assume that FFB SF takes place in halos above a redshift dependent mass threshold derived by \citet{Dekel2023} based on criteria of simultaneously passing a critical volume density, critical surface density, and having Toomre Q less than a critical value. In halos below this mass threshold, the extrapolated results from the \citet{Behroozi2019} abundance matching based UniverseMachine are applied. In galaxies that satisfy these criteria and are in the FFB mode, they assume a fixed star formation efficiency, which is an adjustable parameter of their model. They do not track the build-up of stars through time or self-consistently track the supply of gas that is available for star formation, and the dispersion in galaxy size and formation history at fixed halo mass is not accounted for. Li et al. (in prep) present the results of implementing the FFB criteria self-consistently within a full SAM. 

\citet{Ferrara2023} and \citet{Ferrara2024} present a model that they call the ``Attenuation free model" (AFM). The underlying star formation rate in the model is computed assuming a Kennicutt-Schmidt-like relation, where SFR = $\epsilon_{\rm *}\, f_{\rm b} \, M_{\rm h}/t_{\rm ff}$, where $\epsilon_{\rm *}$ is the star formation efficiency per free fall time, $f_{\rm b}$ is the universal baryon fraction, $M_{\rm h}$ is the halo mass, and $t_{\rm ff}$ is the halo free fall time. Their expression for $\epsilon_{\rm *}$ is a function of the halo circular velocity. Therefore, all redshift dependence in $\epsilon_{\rm *}$ at fixed halo mass arises from the time dependence of the free fall time and the halo circular velocity. The SFR is then converted to UV luminosity assuming a constant factor, including a dust attenuation correction that depends on the sSFR of the galaxy, as described in \S\ref{sec:dust}. \citet{Ferrara2023}, F24 and related works proposed that the evolving impact of dust attenuation with redshift arising from the scaling of sSFR with redshift could provide a complete solution to the puzzle of slower-than-expected evolution of the number density of galaxies to higher redshifts. Our finding that the F24 value of the critical sSFR when applied to our model reproduces the transition from substantial to negligible fractions of galaxies having significant attenuation in the rest-UV at the correct redshift seems non-trivial, given the many differences between our modeling frameworks. It is interesting that applying the F24 sSFR threshold brings our \textbf{cloud-fd=0.1} DMSFE model into good qualitative agreement with the observations at $z\lesssim 12$. However, given that our baseline \textbf{KS} model underpredicts the number of UV-luminous galaxies at higher redshifts $z\gtrsim 12$ even without dust attenuation, this clearly cannot provide a complete solution within the framework of our model (if the current observational results are taken at face value). 

Several studies have previously pointed out the important potential impact of bursty star formation on the number of detected UV-luminous galaxies at high redshift \citep{Mason2023,Shen2023,Sun2023,Kravtsov2024,Gelli2024,Basu2025}. \citet{Shen2023} found that a model with non-evolving star formation efficiency could only reproduce the high redshift $10 \lesssim z \lesssim 14$ UVLF if the 1-$\sigma$ dispersion in M$_{\rm UV}$ at fixed halo mass ($\sigma_{\rm UV}$) was quite high, about 2 (2.5) magnitudes at $z=12$ (14). \citet{Pallottini2023} showed that such a high level of burstyness is not seen in their high resolution SERRA hydrodynamic simulations. \citet{Sun2023} showed that the bursty star formation that is intrinsic to the FIRE-2 cosmological hydrodynamic simulations (which predict that the global integrated SFE barely evolves over the interval $5 \lesssim z \lesssim 12$) helps bring them into better agreement with the observed UVLF at $z\sim 10$--12. \citet{Gelli2024} derived a halo mass dependent, but redshift independent, expression for the amplitude of excursions from an average value of UV magnitude at a given halo mass based on the FIRE-2 simulations. When they combined this with a simple (non-evolving) empirical model for the baseline SFE calibrated to observations at $z=5$, \citet{Gelli2024} showed that this degree of burstyness could not, on its own, explain the discrepancy with observations at $z \gtrsim 11$. Our results are consistent with their conclusion: adding the FIRE-motivated burst amplitude in post-processing to our baseline \textbf{KS} model brings this model into agreement with the observed UVLFs out to $z\sim 12$ but not at higher redshift. We note that the rather steep increase in $\sigma_{\rm UV}$ with decreasing halo mass in the fit provided by \citet{Gelli2024}, combined with the shifting of the most abundant halos to lower masses with increasing redshift that is intrinsic to $\Lambda$CDM, leads to a shallower decline in the number of UV-bright galaxies at $z\gtrsim 9$ when it is applied to our models. Using the high-resolution SPICE simulations \citep{Bhagwat2025}, \citet{Basu2025} found a dependence of $\sigma_{\rm UV}$ on halo mass that is qualitatively similar to the FIRE results, but showed that the amplitude and slope of this halo mass dependence is sensitive to the details of the implementation of stellar feedback in the simulations. 

Large volume hydrodynamic simulations such as IllustrisTNG \citep{Pillepich2018,Nelson2018}, MillenniumTNG \citep{KannanMTNG2023}, THESAN \citep{Kannan2022}, SIMBA and SIMBAEoR \citep{Dave2019,Jones2024}, BlueTides \citep{Feng2016} and FLARES \citep{Wilkins2022} all make predictions that are similar to our baseline \textbf{KS} model; namely, they underproduce UV-luminous galaxies by increasingly large factors at $z\gtrsim 10$, and predict a steeper decline in the comoving number density of UV-bright galaxies at these redshifts than seen in observations \citep{Finkelstein2023,Finkelstein2024,Leung2023,Adams2024}. This is not surprising, as these simulations contain sub-grid models for star formation and supernova-driven winds that are calibrated to reproduce the low integrated global galaxy SFE (i.e. $m_*/(f_{\rm b}M_{\rm h})$ relation) at low redshift. Moreover, these simulations adopt a pressurized effective equation of state \citep[eEOS; see the discussion in][]{SomervilleDave2015}, which is known to lead to less bursty star formation than simulations that at least partially resolve the multi-phase ISM and do not adopt an eEOS \citep{Marinacci2019}. Higher resolution simulations such as FIRE \citep{Hopkins2014,Hopkins2018,Hopkins2023} and FIREBox \citep{Feldmann2023}, SPHYNX \citep{Rosdahl2022}, FirstLight \citep{Ceverino2019,Ceverino2024}, the RAMSES simulations of \citet{Andalman2024} and THESAN-zoom \citep{Kannan2025}, are able to incorporate more physically grounded sub-grid recipes -- but because these simulations are typically zoom-ins or very small volumes, they are limited in their ability to make predictions for the bright end of the UVLF which is the focus of this work. 

\citet{Mauerhofer2025} incorporated information on the cold gas fraction and star formation efficiency from the high resolution {\sc SPHYNX}$^{20}$ simulation \citep{Rosdahl2022} into the {\sc Delphi} semi-analytic model \citep{Dayal2014,Dayal2022,Mauerhofer2023} to make predictions for the evolution of the UVLF from $z\sim 5$--20. They include dust and bursty star formation in their model, where the burstyness arises from their adoption of galaxy properties from the {\sc SPHYNX}$^{20}$ simulations. Due to its small volume, the {\sc SPHYNX}$^{20}$  simulation does not contain the massive halos that are likely the hosts of the UV-luminous galaxies that are observed by JWST at these redshifts. Therefore, \citet{Mauerhofer2025} consider a model variant (eSFE) in which they assign an evolving star formation efficiency to halos in {\sc Delphi} that are more massive than any of the halos in the {\sc SPHYNX}$^{20}$ simulation. Their SFE is defined as the fraction of the cold gas that turns into stars over a 30 Myr timescale, and it is set to have a value of $0.8$ at $z\gtrsim 14$ and to decline as a function of redshift until $z\sim 7$. They also consider an evolving IMF model (eIMF) in which the IMF becomes increasingly top-heavy as a function of both increasing redshift and decreasing metallicity, based on the high resolution hydrodynamical simulations of \citet{Chon2021,Chon2022}. They find that their fiducial model underproduces the bright end of the UVLF at $z\gtrsim 11$, but both their eSFE and eIMF models are consistent with the observations, within the large uncertainties, to $z \sim 14$. 

\citet{Trinca2024} used the Cosmic Archaeology Tool (CAT) semi-analytic model \citep{Trinca2022} to investigate the impact of an evolving IMF as well as the contribution of AGN to the evolution of the UVLF at $4 \lesssim z \lesssim 16$ (CAT predictions at $15 \lesssim z \lesssim 20$ are shown in \citealt{Castellano2025}). Similar to other SAM predictions, \citet{Trinca2024} found that their fiducial model predicts a deficit of UV luminous galaxies at $z\sim 12$--16 of 0.8--1.2 dex, with the deficit increasing towards higher redshift. They found that the contribution of AGN to the UV light is less than 10\% at M$_{\rm UV}<-19$ and $z \gtrsim 10$ in their models. However, the modeling of BH seeding, accretion, and feedback is highly uncertain and these conclusions certainly depend on the details of these modeling choices. In the model variant where black hole seeds
are allowed to grow through short super-Eddington accretion bursts, the AGN
contribution to the UV luminosity function is expected to increase \citep{Trinca2024}.
The CAT model with an evolving IMF, which is also based on the \citet{Chon2021,Chon2022} simulations, boosts the number density of UV luminous galaxies at $z\sim 12$--14 by about 0.6-1 dex. The predicted number density at $M_{\rm UV} \sim -20$ at $z\sim $14--16 is however still a bit low ($\log (dn/dM_{\rm UV}$/[Mpc mag$^{-1}$]) $\sim -6.5$; see \citealp{Trinca2024}, Fig. 12) compared to current observational estimates at this magnitude and redshift ($\log (dn/dM_{\rm UV}$/[Mpc mag$^{-1}$]) $\sim -5.5$ to $-4.5$). 

\section{Conclusions}
\label{sec:conclusions}
In this work, we present a new semi-analytic model with \emph{density modulated star formation efficiency}, motivated by a simple analytic argument that momentum deposition from massive stars must be able to overcome gravity in order to unbind the gas and halt star formation. This leads to a cloud scale star formation efficiency $\epsilon_{*, \rm cl}$ that strongly increases as a function of the cloud surface density, in qualitative agreement with numerical simulations of individual star forming clouds or proto-star-clusters. We combine this with the assumption that the overall galaxy gas radius scales with the virial radius of the halo, which turns out to be consistent with the observed UV sizes of high redshift galaxies, and the assumption that the surface densities of the star forming clouds are comparable to the overall ISM density. Taken together, this implies that the surface densities of star forming clouds in galaxies at $z\gtrsim 10$ are comparable to or even higher than those in super star clusters ($10^4$--$10^5$ \msun), implying that the cloud scale star formation efficiencies are expected to be as high as 80-90\%. We incorporate this new recipe into the Santa Cruz semi-analytic model of galaxy formation, which self-consistently tracks gas inflows and outflows within the backbone of cosmological merger trees extracted from the \gureft\ suite of N-body simulations. We also implemented models for dust attenuation and halo mass dependent bursty star formation in post-processing, inspired by models previously suggested in the literature \citep{Ferrara2024,Gelli2024}. Our main findings are as follows: 

\begin{itemize}
\item The DMSFE model yields number densities of UV luminous galaxies that are up to two orders of magnitude higher at $z\sim 11$--14, and more than four orders of magnitude higher at $z\gtrsim 15$, than our baseline Kennicutt-Schmidt based model (\textbf{KS}). The physically grounded DMSFE models \emph{reproduce or even frequently exceed} the observed galaxy number densities at ultra-high redshift ($z\gtrsim 10$).

\item The DMSFE also predicts higher number densities of massive galaxies at early times $z\sim 6$--12, in better agreement with existing observational estimates of galaxy stellar mass functions than the \textbf{KS} model. 

\item The model predictions are sensitive to the assumed fraction of the total ISM in dense star forming clouds ($f_{\rm dense}$), which we treat as a free parameter.  Although the DMSFE models do produce a more gradual decline in the number density of UV-bright galaxies from $z\sim 6$--14, with any fixed value of $f_{\rm dense}$, this physical process alone as implemented in our models is not able to reproduce the observed \emph{very} shallow decline, especially from $12 \lesssim z \lesssim 14$. However, this should not be considered a failure of the general picture, since it is expected that $f_{\rm dense}$ could have an effective dependence on redshift. 

\item We find that under the ansatz of the \citet{Ferrara2024} model, in which dust is ejected and galaxies become ``attenuation-free'' above a critical specific star formation rate, dust attenuation has a negligible effect on the UVLF at $z\gtrsim 10$ in our models. This is in qualitative agreement with the very blue UV slopes observed in $z\gtrsim 10$ galaxies, and makes the decline of the number density of UV-luminous galaxies more gradual from $6 \lesssim z \lesssim 10$, bringing our model predictions into good agreement with the observations in this redshift range.  

\item Adding bursty star formation with a halo mass dependent amplitude in post-processing to our model leads to a preferential boost in the bright end of the UVLF at higher redshifts, as star formation shifts into less massive halos that have higher burst amplitudes.

\item We suggest that the combined effects of density modulated star formation efficiency, evolving dust attenuation, and evolving bursty star formation, may conspire to yield the surprisingly shallow observed evolution in the number galaxy of UV luminous galaxies at $z\gtrsim 6$. Importantly, in our models, the evolution of these physical processes with cosmic time arises from the indirect effects of the evolution of \emph{physically motivated quantities} (gas surface density, halo potential well depth) due to cosmological structure formation, not inclusion of an ad hoc redshift or time dependence. 

\item Although not (yet) included in our models, the role of an evolving IMF, the contribution of nebular emission, and/or UV light contributed by accreting SMBH remain open questions.

\end{itemize}

In future cycles, JWST has the potential to answer these questions about the relative role of the physical processes that we have highlighted (evolving SFE, dust, starbursts, BH accretion, and the stellar IMF) by searching for higher redshift galaxies at $z \sim 15$--20 with ultra-deep imaging, and obtaining ultra-deep spectroscopy to directly measure outflows, dust attenuation, electron densities and chemical abundances, and constrain the contribution from AGN. 

\section*{Data Availability}
Tabular data for the main results presented in this paper may be downloaded from \url{https://users.flatironinstitute.org/~rsomerville/Data\_Release/DMSFE/}. The catalog level data used in this work will be made available upon request. 

\section*{Acknowledgements}
We warmly thank Avishai Dekel, Zhaozhou Li, Andi Burkert, and Brant Robertson for stimulating discussions. We additionally thank the referee, Raffaella Schneider, for a thorough and constructive report, and Andrea Ferrara for enlightening discussions and useful comments on the manuscript, including pointing out a typographical error in one of the equations. The \gureft\ simulation suite and Santa Cruz Semi-Analytic galaxy formation model was run on the Flatiron Institute computing cluster \textit{rusty}, managed by the Scientific Computing Core (SCC). AY is supported by a Giacconi Fellowship from the Space Telescope Science Institute, which is operated by the Association of Universities for Research in Astronomy, Incorporated, under NASA contract HST NAS5-26555 and JWST NAS5-03127. L.L. gratefully acknowledges the support of the Simons Foundation under grant 965367. The Flatiron Institute is supported by the Simons Foundation. We are grateful to Raffaella Schneider, Brant Robertson, Roberto Maiolino, and Volker Bromm for organizing the Kavli Institute for Theoretical Physics (KITP) program ``Cosmic Origins: The First Billion Years'', and the KITP for hosting this program. This research was supported in part by grant NSF PHY-2309135 to the Kavli Institute for Theoretical Physics (KITP). 




\bibliographystyle{mnras}
\bibliography{ultraz-cloud} 




\appendix

\section{Supplementary figures}
\begin{figure*}
    \includegraphics[width=2\columnwidth]{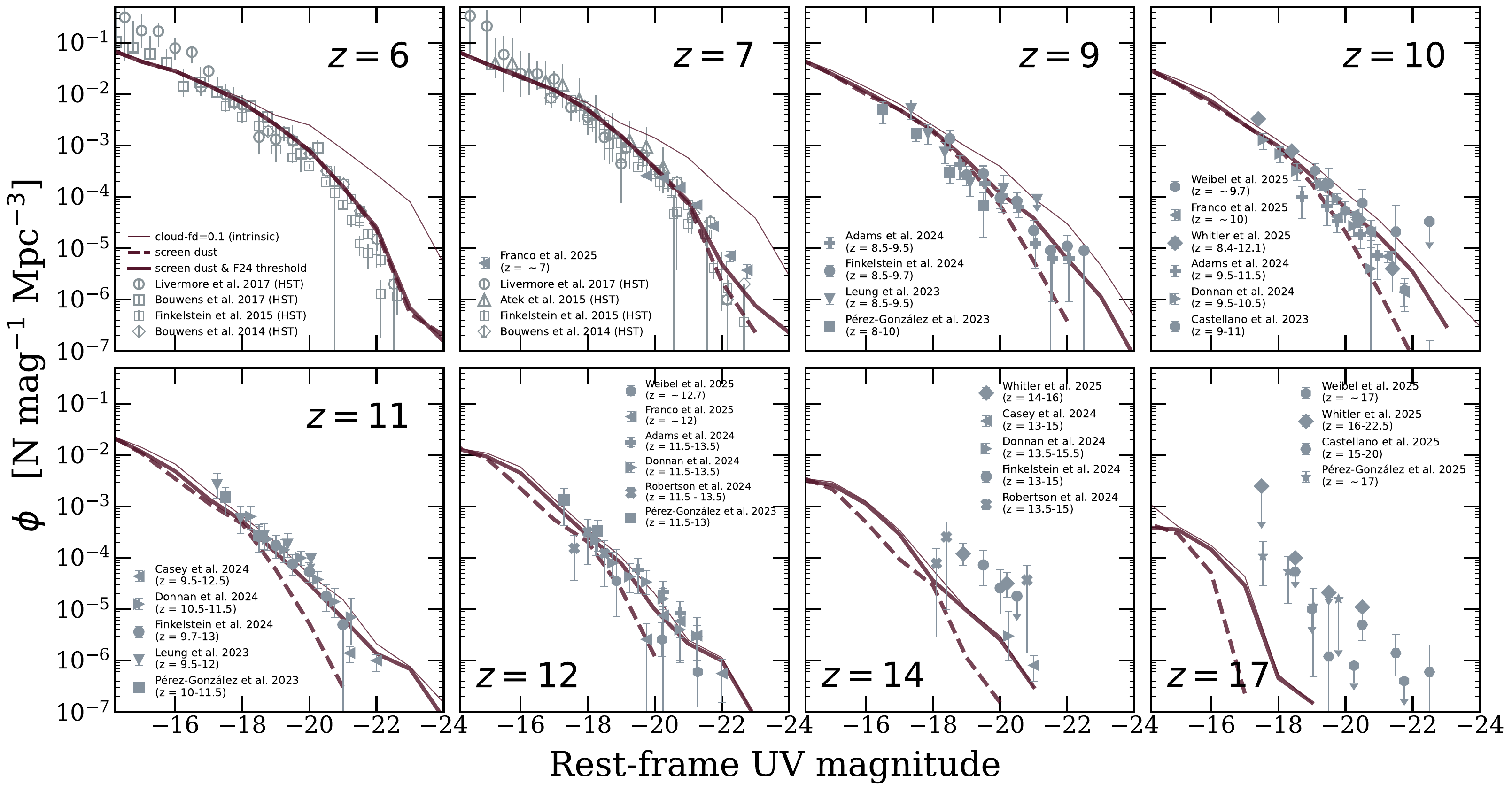}
    \caption{\textbf{Impact of dust}: Rest-UV luminosity functions at redshift $6 < z < 17$. Symbols show a compilation of observational luminosity function estimates, as specified in the figure legend. Light lines show the predictions of the \textbf{cloud-fd=0.1} model without dust attenuation. Dashed lines show the impact of adding dust attenuation using a standard screen model (with no sSFR threshold); heavy solid lines show the predictions of the screen+sSFR$_{\rm crit}$ model.
    }
    \label{fig:UVLF_no_thres}
\end{figure*}

\begin{figure}
    \includegraphics[width=\columnwidth]{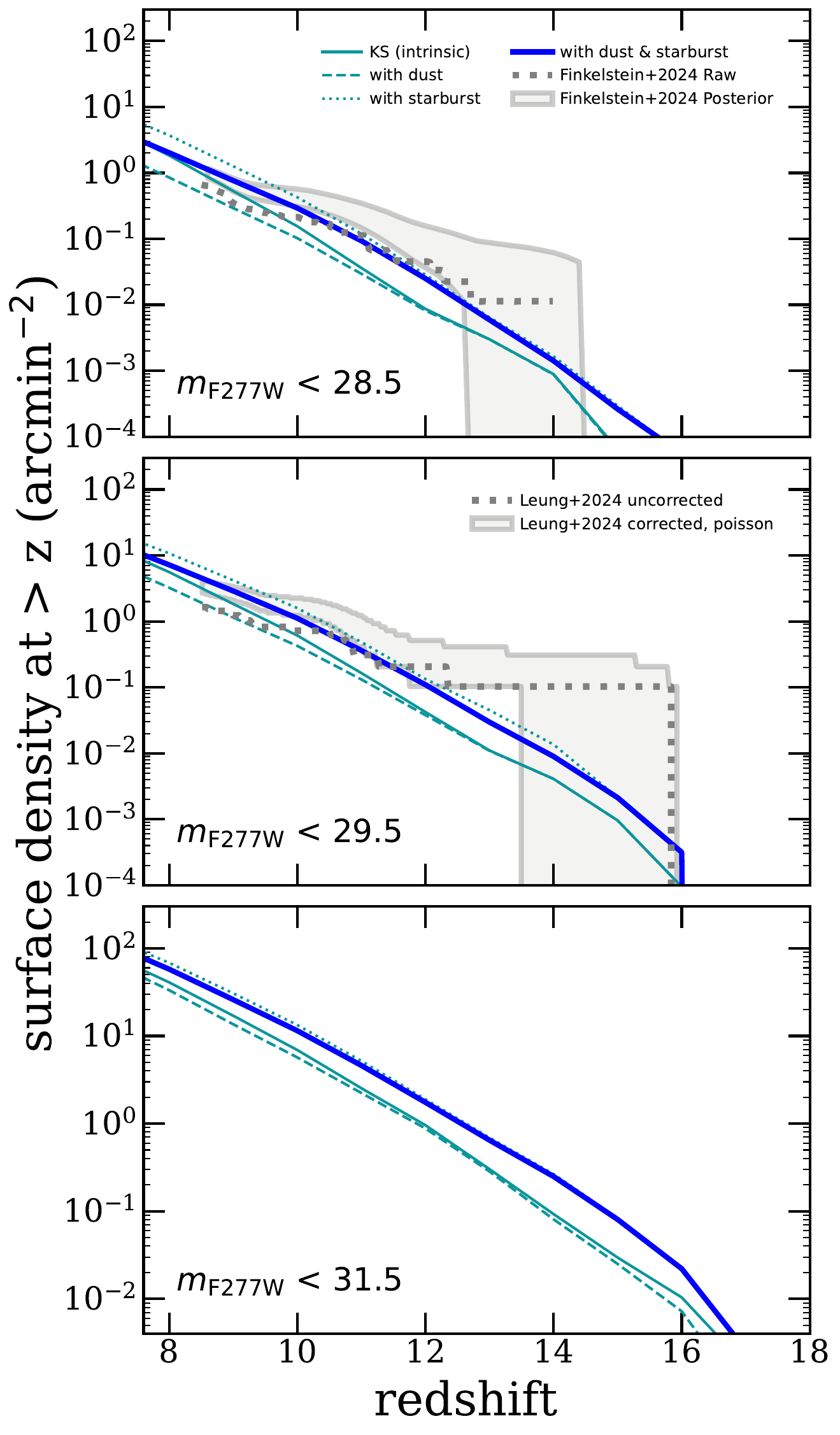}
    \caption{\textbf{KS model with dust and bursts:} Cumulative number counts of galaxies per arcmin$^2$ on the sky with redshifts above the one plotted on the x-axis and apparent magnitude brighter than $m_{\rm 277W}<28.5$ (top), $m_{\rm 277W}<29.5$ (middle), and $m_{\rm 277W}<31.5$ (bottom). The dotted grey lines show the uncorrected number counts from the CEERS (top; \citealp{Finkelstein2024}) and NGDEEP (middle; \citealp{Bagley2023,Leung2023}) surveys and the shaded grey regions show the results from these surveys after a correction for incompleteness; the bottom panel represents a possible future ultra-deep JWST survey. Lighter cyan lines show the models without dust and without bursts. Dashed lines show the models with the inclusion of dust, but no bursts. Dotted lines show the impact of including bursts, but not dust. Dark blue solid lines show the impact of including bursts and dust. 
    }
    \label{fig:counts_dust_n_burst_KS}
\end{figure}

\begin{figure}
    \includegraphics[width=\columnwidth]{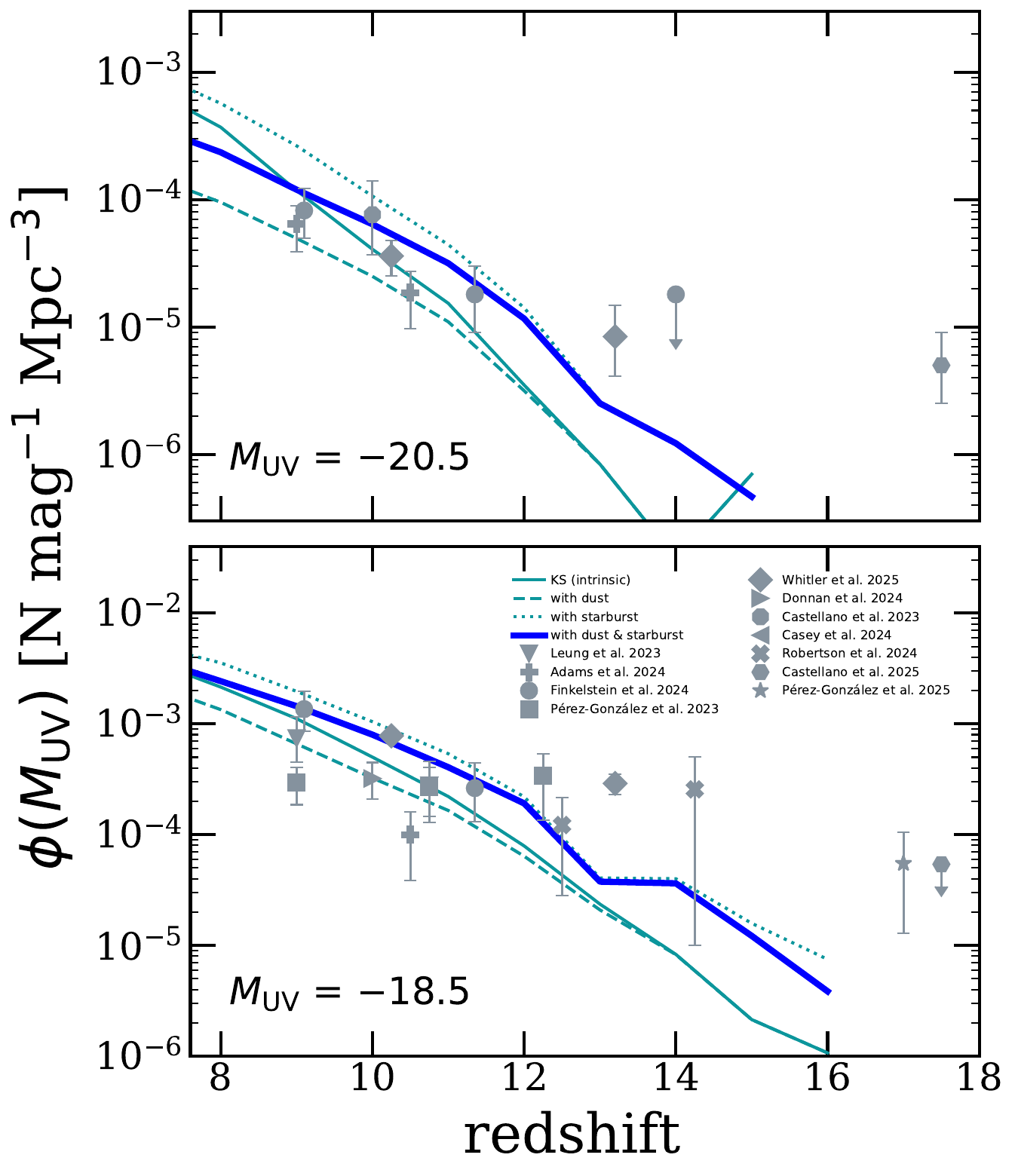}
    \caption{The comoving number density of galaxies with $M_{\rm UV}=-20.5$ (top) and $M_{\rm UV}=-18.5$ (bottom) as a function of redshift. Symbols show observational estimates, and predictions from the \textbf{KS} model with different permutations of dust and bursts included in post-processing are shown with the same line types as in Fig.~\ref{fig:counts_dust_n_burst_KS}.  
    }
    \label{fig:nvsz_KS}
\end{figure}

In this appendix, we show a few supplemental figures. Fig.~\ref{fig:UVLF_no_thres} shows the UVLF for the \textbf{cloud-fd=0.1} model with a standard screen dust model as well as the screen+sSFR$_{\rm crit}$ dust model shown in the main body of the paper. Fig.~\ref{fig:counts_dust_n_burst_KS} and Fig.~\ref{fig:nvsz_KS} show the impact of dust and starbursts on our baseline \textbf{KS} model. 

\bsp	
\label{lastpage}
\end{document}